\documentclass[10pt, conference, compsoc]{IEEEtran}

\IEEEoverridecommandlockouts
\ifCLASSINFOpdf
\else
\fi

\usepackage{color}
\usepackage{pstricks}
\usepackage{enumerate}

\usepackage{tikz}
\usepackage{tikz-cd}

\usepackage{amssymb}
\usepackage{amstext}
\usepackage{amsmath}
\usepackage{stmaryrd}
\usepackage{bbold}

\renewcommand{\to}{\rightarrow}

\newcommand{\tto}[1]{\xrightarrow{#1}}
\newcommand{\oot}[1]{\xleftarrow{#1}}

\newcommand{\epi}{\twoheadrightarrow}
\newcommand{\eepi}[1]{\xrightarrow{#1}\hspace{-1.3ex}\rightarrow}

\newcommand{\WP}{\mbox{\Large $\wp$}}

\newcommand{\Set}{{\sf Set}}
\newcommand{\Pfn}{{\sf Pfn}}

\newcommand{\Rel}{{\sf Rel}}

\newcommand{\id}{{\rm id}}

\newcommand{\CCC}{{\cal C}}

\newcommand{\GGG}{{\cal G}}

\newcommand{\MMM}{{\cal M}}

\newcommand{\OOO}{{\cal O}}

\newcommand{\TTT}{{\cal T}}

\renewcommand{\Bbb}{\mathbb}

\newcommand{\MMm}{{\Bbb M}}
\newcommand{\NNn}{{\Bbb N}}

\newcommand{\PPp}{{\Bbb P}}

\mathcode`\<="4268 
\mathcode`\>="5269 
\mathchardef\gt="313E 
\mathchardef\lt="313C 

 \def\pushright#1{{
    \parfillskip=0pt            
    \widowpenalty=10000         
    \displaywidowpenalty=10000  
    \finalhyphendemerits=0      
    \leavevmode                 
    \unskip                     
    \nobreak                    
    \hfil                       
    \penalty50                  
    \hskip.2em                  
    \null                       
    \hfill                      
    {#1}                        
    \par}}                      

 \def\qed{\pushright{$\square$}\penalty-700 \smallskip}

\newenvironment{prf}[1]{\begin{trivlist} \item[{\it ~Proof}#1.]}%
{\qed\end{trivlist}}

\newcommand{\be}[1]{\begin{#1}}

\newcommand{\ee}[1]{\end{#1}}
\newcommand{\beq}{\begin{equation}}
\newcommand{\eeq}{\end{equation}}
\newcommand{\ba}[1]{\begin{array}{#1}}
\newcommand{\ea}{\end{array}}
\newcommand{\bea}{\begin{eqnarray}}
\newcommand{\eea}{\end{eqnarray}}
\newcommand{\bear}{\begin{eqnarray*}}
\newcommand{\eear}{\end{eqnarray*}}
\newcommand{\bpr}{\begin{prf}{}}
\newcommand{\epr}{\end{prf}}
\newcommand{\bprf}[1]{\begin{prf}{#1}}
\newcommand{\eprf}{\end{prf}}

\newtheorem{thm}{Theorem}[section]

\newtheorem{cond}{}[thm]

\newtheorem{prenumb}[thm]{\hspace{-1ex}}

\newcommand{\comp}[2]{{#1}\hspace{.2ex} ; {#2}}
\newcommand{\compp}{(\, ;)}
\newcommand{\Nn}{\PP}

\newcommand{\UK}{{\sf u}}
\newcommand{\VK}{{\sf w}}
\newcommand{\TK}{{\sf t}}
\newcommand{\SK}{{\sf s}}

\newcommand{\CX}{{\sf c}}
\newcommand{\GK}{\kappa}

\newcommand{\enco}[1]{\left\ulcorner{#1}\right\urcorner}

\newcommand{\mnd}{\varrho}

\newcommand{\cmn}{\delta}
\newcommand{\cun}{{\scriptstyle \top}}

\newcommand{\PP}{\PPp}
\newcommand{\MMMM}{\MMM}
\newcommand{\MMmm}{\MMm}
\newcommand{\doublone}{\mathbb{1}}
\newcommand{\Bits}{\mathbb{2}}

\newcommand{\NNnn}{\overline\NNn}

\newcommand{\restr}{\!\!\restriction}

\newcommand{\grad}[1]{\left\|{#1}\right\|}

\newcommand{\opls}{\oplus}

\newcommand{\initstate}{q}
\newcommand{\initdata}{a}
\newcommand{\uu}{w}
\newcommand{\KT}{T}

\begin{document}
\title{
Monoidal computer II:\\ 
Normal complexity by string diagrams}

\author{\IEEEauthorblockN{Dusko Pavlovic}
\IEEEauthorblockA{University of Hawaii\\
Email: dusko@hawaii.edu}
}

\IEEEspecialpapernotice{(Extended abstract)}

\maketitle

\begin{abstract}
In \emph{Monoidal Computer I}, we introduced a categorical model of computation where the formal reasoning about computability was supported by the simple and popular diagrammatic language of \emph{string diagrams}. In the present paper, we refine and extend that model of computation to support a formal complexity theory as well. This formalization brings to the foreground the concept of \emph{normal}\/ complexity measures, which allow decompositions akin to Kleene's normal form. Such measures turn out to be just those where evaluating the complexity of a program does not require substantially more resources than evaluating the program itself. The usual time and space complexity are thus normal measures, whereas the average and the randomized complexity measures are not. While the measures that are not normal provide important design time information about algorithms, and for theoretical analyses, normal measures can also be used at run time, as practical tools of computation, e.g. to set the bounds for hypothesis testing, inductive inference and algorithmic learning.  
%
\end{abstract}

\begin{IEEEkeywords}
categorical models and logics,
foundations of computability, logical aspects of computational complexity, semantics of computation
\end{IEEEkeywords}

%
\IEEEpeerreviewmaketitle

\section{Introduction}
\subsubsection*{Motivation} This is an old fashioned paper, inspired by Blum's \emph{"Machine Independent Complexity Theory"}\/ \cite{BlumM:axioms}, Levin's and Meyer's work on the \emph{"Fundamental Theorem of Complexity"}  \cite{LevinL:fundamental,MeyerA:speedup}, and a host of such papers --- mostly from the 1970s. Why is it worthwhile to go back to that work, when theoretical computer science went in a different direction?

Why is it reasonable that theoretical computer science still  works with tapes, and still programs its abstract machines in low level machine languages, long after the real computers stopped using tapes, and when the real programmers only use machine languages to program firmware?  There are surely some reasonable and convincing answers to this question, and there are probably some good reasons why the theoreticians like to use low level models. But as theoretical computer science is becoming more and more practical, the practical tasks are emerging where high level models and machine independent reasoning and programming are becoming necessary. One such task \cite{PavlovicD:NSPW11} led me on the path towards monoidal computer. 
The task was to measure the hardness of deriving an attack algorithm on a given system with known vulnerabilities. This task is easily formalized, but requires measuring the complexity of algorithm transformations. There are high level programming languages convenient for programming program transformations, but there are no research tools for studying the complexity of such programs. Wondering why, I turned to a tool that seemed useful for understanding high level programming languages: \emph{category theory}. 

So my defense for the strange and demanding concoction of the formalisms that I offer here is very ambitious: I am hoping that it will get us a step closer to a high level language for reasoning about computability, complexity and cryptography, by reusing some ideas and structures that evolved in such languages for reasoning about software, systems, and even about quantum computing. If this turns out to be a completely wrong direction, then we shall at least gain some insight why the problem of complexity is so different from pretty much everything else.

\subsubsection*{Idea} Intuitively, computation is often viewed as a straightforward process.  A computer is given a program $F$, and it is set into a  configuration $<\initstate , \initdata >$, where $\initstate $ is the initial state, and $\initdata $ are the input data. The computer then searches for an instruction of the program $F$ that is enabled by the configuration $<\initstate , \initdata >$, and if it finds such an instruction, it executes it, thereby changing the state to $q_1$ and the data to $a_1$. Then it searches for an instruction enabled by the configuration $<q_1, a_1>$, which leads to a configuration $<q_2, a_2>$, and so on. The computer thus builds an \emph{execution trace}, which can be viewed as an expression in the form
\beq\label{trace}
F   : <\initstate ,\initdata > \to <q_1,a_1>\to \cdots \to <q_i,a_i>\to \cdots
\eeq
If the trace reaches a configuration $<q_n,a_n>$ where no  instructions of the program $F$ are enabled, then the computation terminates.  The output of the computation can then be found among the data $a_n$, say as the part that is stored on a designated output tape. 

Kleene's Normal Form Theorem \cite{KleeneSC1936} formalizes this view of computation mathematically. It says that every computable function $f:A \to B$, implemented by some program $F$, can be reduced to the \emph{normal form}
\bea\label{normform}
f(a) & = & \uu\big( \mu x.\ \KT(F,a,x)\big)
\eea
where $\uu:\NNn\to \NNn$ and $\KT:\NNn^3 \to \{0,1\}$ are some primitive recursive functions, explained below, and $\mu x:\{0,1\}^\NNn \to \NNn$ is the search operator.  
What does \eqref{normform} mean? The idea is that any execution trace \eqref{trace}, as soon as it is finite, and thus denotes a terminating computation, can be encoded by a unique natural number $x$. Kleene constructed a primitive recursive predicate $\KT(F,a,x)$ that tests whether $x$ encodes the trace \eqref{trace} of an execution of the program $F$  on data $a$. 
If the search $\mu x.\KT(F,a,x)$ finds such a trace $x$, then the primitive recursive function $\uu$ extracts from its final configuration the output of the computation of $F$ on $a$. Formula \eqref{normform} thus tells that every computation can be reduced to a single search $\mu x$, precomposed with a primitive recursive predicate $T(F,a,x)$ which tests that $x$ is the trace of the program $F$ on the input $a$, and postcomposed with a primitive recursive function $\uu$ which extracts the outputs from the traces.

Note, however, that besides the program, the input and the output, the trace \eqref{trace} also carries the information how many steps did the computation take, and what was the largest memory area that it occupied. So if we replace the primitive recursive function $\uu$ with some other functions, we can compute the time and the space complexity. That is the idea that we pursue in this paper. It arises from the observation that the suitable encodings of execution traces are not only complexity measures themselves, but that they are universal among a natural family of complexity measures, which we call normal. We present them as a categorical structure, and calculate with them using string diagrams.

\subsubsection*{Outline of the paper} In Sec.~\ref{Sec-Prelim}, we spell out the minimal preliminaries that fit in this paper. In Sec.~\ref{Sec-MonCom}, we review the structure of monoidal computer. Sec.~\ref{Sec-Grad} introduces the structure of \emph{graded monoidal computer}. The grades implement the execution traces categorically. In Sec.~\ref{Sec-Norm}, we spell out the normalization in graded monoidal computer. In Sec.~\ref{Sec-CX} we define and characterize normal complexity measures. 

\subsubsection*{Related work}
While computability and complexity theorists seldom felt a need to learn about categories, there is a rich tradition of categorical research in computability theory, starting from one of the founders of category theory and his students \cite{eilenberg-elgot,heller-dipaola}, through the extensive categorical investigations of realizability \cite{HylandJ:efft,BirkedalL,HofstraW13}, to the recent work on Turing categories \cite{Cockett-Hofstra}, and on a monoidal structure of Turing machines \cite{BarthaM:monoidal-TM}. This recent work has, of course, interesting correlations with basic monoidal computer, but also substantial differences, arising from the different goals. The closest in spirit to the present work seems \cite{AspertiA:rice}, also drawing its structural content from abstract complexity theory.

\section{Preliminaries}\label{Sec-Prelim}
A monoidal computer will be a \emph{symmetric monoidal category}, with some additional structure. As a matter of convenience, and with no loss of generality, we assume that it is a \emph{strict}\/ monoidal category. The reader familiar with these concepts may wish to skip to the next section. For the casual reader unfamiliar with these concepts, we attempt to provide enough intuitions to understand the presented ideas. The reader interested to learn more about monoidal categories should consult one of many textbooks, e.g. \cite{MacLaneS:CWM,KellyM:Enriched}. 

\subsubsection*{Monoidal categories} 
Intuitively, a monoidal category is a category $\CCC$ together with a functorial monoid structure 
$$\CCC\times \CCC \tto \otimes \CCC \oot I \doublone$$ When $\CCC$ is a monoidal \emph{computer}, then we think of its objects $A, B, \ldots \in |\CCC|$ as datatypes, and of its morphisms, $f, g\ldots\in \CCC(A,B)$ as computations. The tensor product $A\otimes P\tto{f \otimes t} B\otimes Q$ then captures the parallel composition of the computations $A\tto f B$ and $P\tto t Q$, whereas the categorical composition $A\tto{\comp f g} C$  is the sequential composition of $A\tto f B$ and $B\tto g C$. 

With no loss of generality, we assume that the tensors are \emph{strictly}\/ associative and unitary, and thus treat the objects $A\otimes (B\otimes C)$ and $(A\otimes B)\otimes C$ as the same, and do not distinguish $A\otimes I$ and $I\otimes A$ from $A$. This allows us to elide many parentheses and natural coherences \cite[Sec.~VII.2]{Joyal-Street:geometry,MacLaneS:CWM}. Note, however, that the isomorphisms $A\otimes B \tto \varsigma B\otimes A$ cannot be eliminated without causing a degeneracy. 

\subsubsection*{Notations} When no confusion seems likely, we write 
\begin{itemize}
\item $AB$ instead of $A\otimes B$
\item $\CCC(X)$ instead of $\CCC(I,X)$
\end{itemize}

\subsubsection*{String diagrams} A salient feature of monoidal categories is that the algebraic laws of the monoidal structure correspond precisely and conveniently to the geometric laws of \emph{string diagrams}, formalized in \cite{Joyal-Street:geometry}, but going back to \cite{Penrose}. See also \cite{selinger2011survey} for survey. 
A string diagram usually consists of polygons or circles linked by strings. In a monoidal computer, the polygons represent computations, whereas the strings represent data types, or the channels through which the data of the corresponding types flow. String diagrams thus display the data flows throught composite computations. The reason why string diagrams are convenient for this is that the two program operations that usually generate data flows, the sequential composition $\comp f g$ and the parallel composition $f\otimes t$, precisely correspond to the two geometric operations that generate string diagrams: one is the operation of connecting the polygons $A\tto f B$ and $B\tto g C$ by the string $B$, whereas the other one puts the polygons $A\tto f B$ and $P\tto t Q$ next to each other without connecting them. 
The associativity of these geometric operations then imposes the associativity law on the corresponding operations on computations. The identity morphism $\id_A$, as the unit of the sequential composition, can be viewed as the channel of type $A$, and can thus be presented as the string $A$ itself, or as an "invisible polygon" freely moved along the string $A$. The unit type $I$ can be similarly presented as an "invisible string", freely added and removed to string diagrams. The algebraic laws of the monoidal structure are thus captured by the geometric properties of the string diagrams. The string crossings correspond to the symmetries $A\otimes B \tto \varsigma  B\otimes A$. 

\subsubsection*{Data services} 
We call \emph{data service}\/ the monoidal structure that allows passing the data around. In computer programs and in mathematical formulas, the data are usually passed around using variables. They allow copying and propagating the data values where they are needed, or deleting them when they are not needed. The basic features of a variable are thus that it can be freely copied or deleted. The basic data services over a type $A$ in a monoidal category $\CCC$ are 
\begin{itemize}
\item the \emph{copying}\/ operation $A\tto \cmn A\otimes A$, and 
\item the \emph{deleting}\/ operation $A\tto \cun I$, 
\end{itemize}
which together form a \emph{comonoid}, i.e. 
satisfy the equations
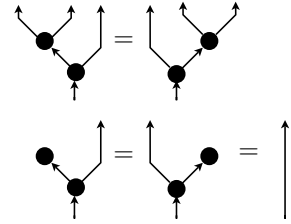
\begin{gather}
\comp {\cmn}{(\cmn \otimes A)}  = \comp{\cmn}{(A\otimes \cmn)} \hspace{3.3em} \label{eq-assoc}\\
\comp{\cmn}{(\cun \otimes A)}  = \comp{\cmn}{(A\otimes \cun)} = \id_A
\end{gather}
\begin{center}
\newcommand{\comonoidd}{\cmn}
\newcommand{\comonunn}{\cun}
\newcommand{\comontwo}{\cmn}
\def\JPicScale{.7}
\ifx\JPicScale\undefined\def\JPicScale{1}\fi
\psset{unit=\JPicScale mm}
\psset{linewidth=0.3,dotsep=1,hatchwidth=0.3,hatchsep=1.5,shadowsize=1,dimen=middle}
\psset{dotsize=0.7 2.5,dotscale=1 1,fillcolor=black}
\psset{arrowsize=1 2,arrowlength=1,arrowinset=0.25,tbarsize=0.7 5,bracketlength=0.15,rbracketlength=0.15}
\begin{pspicture}(0,0)(50,40.62)
\rput(43.12,11.88){$=$}
\rput(19.38,33.12){$=$}
\rput{0}(35.63,33.12){\psellipse[fillstyle=solid](0,0)(1.56,-1.56)}
\psline{<-}(40.62,40.62)
(40.62,40)
(40.63,38.12)(36.88,34.37)
\psline{<-}(30.62,40.62)
(30.62,40)
(30.63,38.12)(34.38,34.37)
\rput{0}(29.38,26.87){\psellipse[fillstyle=solid](0,0)(1.56,-1.56)}
\psline{<-}(5.62,31.88)
(10.62,26.88)
(9.38,26.88)(10,26.88)
\psline{<-}(24.37,40)
(24.38,33.75)
(24.38,31.87)(28.13,28.12)
\rput{0}(4.38,33.12){\psellipse[fillstyle=solid](0,0)(1.56,-1.56)}
\rput{0}(10.31,27.19){\psellipse[fillstyle=solid](0,0)(1.56,-1.56)}
\psline{<-}(15,40)
(15,37.5)
(15,31.87)(11.25,28.12)
\pscustom[]{\psline{<-}(34.38,31.88)(29.38,26.88)
\psbezier(29.38,26.88)(29.38,26.88)(29.38,26.88)
\psbezier{-}(29.38,26.88)(29.38,26.88)(29.38,26.88)
}
\psline{<-}(9.37,40)
(9.37,39.37)
(9.38,37.5)(5.63,33.75)
\psline{<-}(-0.63,40)
(-0.63,39.37)
(-0.62,37.5)(3.13,33.75)
\rput(19.38,11.25){$=$}
\rput{0}(35.63,11.25){\psellipse[fillstyle=solid](0,0)(1.56,-1.56)}
\rput{0}(29.38,5){\psellipse[fillstyle=solid](0,0)(1.56,-1.56)}
\psline{<-}(5.62,10)
(10.62,5)
(9.38,5)(10,5)
\psline{<-}(29.38,3.75)
(29.38,0)
(29.38,-0.62)(29.38,0)
\psline{<-}(24.37,18.12)
(24.38,11.87)
(24.38,10)(28.13,6.25)
\rput{0}(4.38,11.25){\psellipse[fillstyle=solid](0,0)(1.56,-1.56)}
\rput{0}(10.31,5.31){\psellipse[fillstyle=solid](0,0)(1.56,-1.56)}
\psline{<-}(15,18.12)
(15,15.62)
(15,10)(11.25,6.25)
\psline{<-}(10,3.75)
(10,0)
(10,-0.62)(10,0)
\pscustom[]{\psline{<-}(34.38,10)(29.38,5)
\psbezier(29.38,5)(29.38,5)(29.38,5)
\psbezier{-}(29.38,5)(29.38,5)(29.38,5)
}
\psline{<-}(29.38,25.62)
(29.38,22.5)
(29.38,21.88)(29.38,22.5)
\psline{<-}(10,25.62)
(10,22.5)
(10,21.88)(10,22.5)
\psline{<-}(50,18.12)
(50,0)
(50,-0.62)(50,0)
\end{pspicture}

\end{center}
The correspondence between the variables and the comonoids was formalized and explained in \cite{PavlovicD:MSCS97,PavlovicD:Qabs12}. 
The associativity and the unit of the copying operation allow defining the unique $n$-ary copying operations $A\tto\cmn A^{\otimes n}$, for all $n\geq 0$. The tensor products $\otimes$ in $\CCC$ are the cartesian products $\times$ if and only if every $A$ in $\CCC$ carries a canonical comonoid $A\times A\oot \cmn A \tto \cun \doublone$, where $\doublone$ is the final object of $\CCC$, and all morphisms of $\CCC$ are comonoid homomorphisms, or equivalently, the families $A\tto \cmn A\times A$ and $A\tto \cun \doublone$ are natural. Cartesian categories are thus just  the categories with natural copying and deleting operations.

But besides copying and deleting, we often also need a third data service:
\begin{itemize}
\item the \emph{comparison}\/ operation $A\otimes A \tto \mnd A$
\end{itemize}
which is required to be associative, in the sense dual to \eqref{eq-assoc}, and thus makes $A$ into a \emph{semigroup}. Its associativity allows defining the unique $n$-ary comparisons $A^{\otimes n}\tto\mnd A$, for $n\geq 1$. The copying and the comparison operations are further required to satisfy the \emph{data distribution}\/ (or \emph{Frobenius}\/ \cite{Carboni-Walters}) conditions
{\small \[
\comp {(\cmn\otimes A)}{(A\otimes \mnd)}= \comp \mnd \cmn  = \comp{(A\otimes \cmn)}{(\mnd\otimes A)} \qquad \comp \cmn \mnd = \id \qquad
\]}
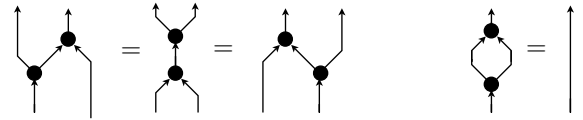
\begin{center}
\newcommand{\eqls}{=}
\newcommand{\comonoid}{\scriptstyle \cmn}
\newcommand{\monoid}{\scriptstyle \mnd}
\def\JPicScale{.6}
\ifx\JPicScale\undefined\def\JPicScale{1}\fi
\psset{unit=\JPicScale mm}
\psset{linewidth=0.3,dotsep=1,hatchwidth=0.3,hatchsep=1.5,shadowsize=1,dimen=middle}
\psset{dotsize=0.7 2.5,dotscale=1 1,fillcolor=black}
\psset{arrowsize=1 2,arrowlength=1,arrowinset=0.25,tbarsize=0.7 5,bracketlength=0.15,rbracketlength=0.15}
\begin{pspicture}(0,0)(126.25,23.75)
\rput(45.75,17.75){}
\rput(28.88,12.75){$\eqls$}
\rput(49.38,13.12){$\eqls$}
\rput(118.25,8.38){}
\rput(118.75,13.12){$\eqls$}
\rput{0}(7.5,8.12){\psellipse[fillstyle=solid](0,0)(1.56,-1.56)}
\psline{->}(7.5,8.12)
(3.75,11.88)
(3.75,20)(3.75,23.12)
\rput{0}(38.75,16.25){\psellipse[fillstyle=solid](0,0)(1.56,-1.56)}
\psline{->}(38.75,16.25)
(35.62,19.38)
(34.38,20.62)(34.38,23.75)
\psline{->}(38.75,11.88)
(38.75,9.38)
(38.75,11.88)(38.75,15)
\psline{->}(39.38,16.88)
(43.12,20.62)
(43.12,21.88)(43.12,23.75)
\rput{0}(38.75,8.12){\psellipse[fillstyle=solid](0,0)(1.56,-1.56)}
\psline{->}(34.38,-0.62)
(34.38,1.25)
(34.38,3.75)(37.5,6.88)
\psline{->}(43.75,-0.62)
(43.75,1.25)
(43.75,3.12)(40,6.88)
\rput{0}(70.94,8.12){\psellipse[fillstyle=solid](0,0)(1.56,-1.56)}
\rput{0}(63.12,15.62){\psellipse[fillstyle=solid](0,0)(1.56,-1.56)}
\psline{->}(70.62,8.12)
(75.62,13.12)
(75.62,17.5)(75.62,22.5)
\psline{->}(70.62,1.25)
(70.62,-0.62)
(70.62,1.25)(70.62,6.88)
\pscustom[]{\psbezier{-}(70.62,8.12)(70.62,8.12)(70.62,8.12)(70.62,8.12)
\psbezier(70.62,8.12)(70.62,8.12)(70.62,8.12)
\psline{->}(70.62,8.12)(64.38,14.38)
}
\psline{->}(63.12,16.88)
(63.12,14.38)
(63.12,16.88)(63.12,22.5)
\psline{->}(58.12,-0.62)
(58.12,4.38)
(58.12,10.62)(61.88,14.38)
\rput(116.38,9.62){}
\rput{0}(108.75,5.62){\psellipse[fillstyle=solid](0,0)(1.56,-1.56)}
\pscustom[]{\psline(108.75,5.62)(104.38,10)
\psbezier(104.38,10)(104.38,10)(104.38,10)
\psline(104.38,10)(104.38,13.12)
}
\psline{->}(108.75,1.25)
(108.75,-0.62)
(108.75,1.25)(108.75,4.38)
\psline(109.38,6.25)
(113.12,10)
(113.12,11.25)(113.12,13.12)
\rput{0}(108.75,17.5){\psellipse[fillstyle=solid](0,0)(1.56,-1.56)}
\psline{->}(104.38,11.25)
(104.38,10.62)
(104.38,13.12)(107.5,16.25)
\psline{->}(113.12,10.62)
(113.12,11.25)
(113.12,13.12)(110,16.25)
\psline{->}(108.75,19.38)
(108.75,16.88)
(108.75,19.38)(108.75,22.5)
\psline{->}(126.25,2.5)
(126.25,-0.62)
(126.25,2.5)(126.25,23.12)
\rput{0}(15,15.62){\psellipse[fillstyle=solid](0,0)(1.56,-1.56)}
\psline{->}(6.88,7.5)
(8.12,8.75)
(10,10.62)(13.75,14.38)
\psline{->}(15,16.88)
(15,14.38)
(15,16.88)(15,22.5)
\psline{->}(20,-1.88)
(20,8.75)
(20,10.62)(16.25,14.38)
\psline{->}(7.5,1.25)
(7.5,-0.62)
(7.5,1.25)(7.5,6.88)
\end{pspicture}

\end{center}
These conditions allow factoring any morphism $A^{\otimes m}\to A^{\otimes n}$ generated by $\cmn$ and $\mnd$ into the \emph{spider}\/ form $A^{\otimes m}\tto\mnd A \tto \cmn A^{\otimes n}$. In summary,

\be{defn}
A \emph{data service}\/ over an object $A$ of a monoidal category $\CCC$ consists of
\begin{itemize}
\item the \emph{copying}\/ operation $A\tto \cmn A \otimes A$, 
\item the \emph{deleting}\/ operation $A\tto \cun I$, and
\item the \emph{comparison}\/ operation $A \otimes A \tto \mnd A$
\end{itemize}
such that
\begin{itemize}
\item $\cmn$ and $\mnd$ are associative,
\item $\cun$ is the unit of $\cmn$,
\item $\cmn$ and $\mnd$ satisfy the data distribution conditions.
\end{itemize}
%
\ee{defn}

\subsubsection*{Examples and non-examples of data services} Nontrivial cartesian categories do not support comparisons. However, the cartesian structures $A\times A \oot \cmn A\tto \cun \doublone$ from the category $\Set$ of sets and functions also live in the monoidal categories $\Rel$ of sets and relations and $\Pfn$ of sets and partial functions. These categories are not cartesian because the singleton maps are only natural with respect to the total relations, whereas the diagonals are only natural with respect to the single-valued relations (i.e. partial functions). Both $\Rel$ and $\Pfn$ allow comparisons: a comparison $A\times A\tto \mnd A$ is the converse of the diagonal, i.e. $\mnd(x,x) = x$, and $\mnd(x,y)$ remains undefined if $x\neq y$. So the cartesian structure of $\Set$ provides the standard data services in $\Rel$ and $\Pfn$. In addition, $\Rel$ also admits many nonstandard data services, that do not come from the cartesian structure. This was analyzed in \cite{PavlovicD:QI09,PavlovicD:Qabs12}. Indeed, a Any abelian group (or groupoid) structure $A\times A \tto + A$ can be used as the comparison operation, with the corresponding copying operation $A \tto \cmn A\times A$ relating each $x\in A$ with all pairs $<y,z>\in A\times A$ such that $x=y+z$. The deletion operation $A\tto \cun \doublone$ relates the unit of the group $A$ with the only element of $\doublone$.

\subsubsection*{Functions}
A morphism $f\in \CCC(A,B)$ is called \emph{function}\/ if it is a comonoid homomorphism with respect to the data services on $A$ and $B$, i.e. if it satisfies the following equations
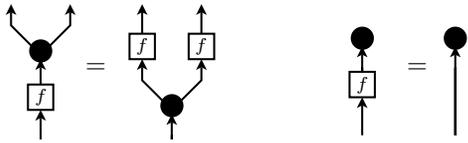
\beq \label{eq-comon-hom}
\qquad\comp f {\cmn_B} \ =\  \comp {\cmn_A}{(f\otimes f)}\qquad \quad\comp f {\cun_B}\ =\ \cun_A
\eeq

\begin{center}
\newcommand{\monoidd}{\cmn}
\newcommand{\fun}{\scriptstyle f}
\newcommand{\One}{A}
\newcommand{\Two}{B}
\newcommand{\delete}{\cun}
\def\JPicScale{.9}
\ifx\JPicScale\undefined\def\JPicScale{1}\fi
\psset{unit=\JPicScale mm}
\psset{linewidth=0.3,dotsep=1,hatchwidth=0.3,hatchsep=1.5,shadowsize=1,dimen=middle}
\psset{dotsize=0.7 2.5,dotscale=1 1,fillcolor=black}
\psset{arrowsize=1 2,arrowlength=1,arrowinset=0.25,tbarsize=0.7 5,bracketlength=0.15,rbracketlength=0.15}
\begin{pspicture}(0,0)(66.56,19.38)
\rput(11.88,10){$=$}
\pspolygon[](1.88,7.5)(5.62,7.5)(5.62,3.75)(1.88,3.75)
\pspolygon[](16.88,15.01)(20.63,15.01)(20.63,11.25)(16.88,11.25)
\pspolygon[](25.62,15.01)(29.39,15.01)(29.39,11.25)(25.62,11.25)
\rput(3.75,5.62){$\fun$}
\rput(18.75,13.12){$\fun$}
\rput(27.5,13.12){$\fun$}
\rput(59.38,10){$=$}
\pspolygon[](49.38,9.38)(53.12,9.38)(53.12,5.62)(49.38,5.62)
\rput(51.26,7.5){$\fun$}
\rput{0}(3.75,12.5){\psellipse[fillstyle=solid](0,0)(1.56,-1.56)}
\pscustom[]{\psline{-}(3.75,8.12)(3.75,7.5)
\psbezier(3.75,7.5)(3.75,7.5)(3.75,7.5)
\psline{->}(3.75,7.5)(3.75,11.25)
}
\pscustom[]{\psline{-}(3.75,0)(3.75,-0.62)
\psbezier(3.75,-0.62)(3.75,-0.62)(3.75,-0.62)
\psline{->}(3.75,-0.62)(3.75,3.75)
}
\psline{->}(3.75,11.88)
(0.62,15)
(-0.62,16.25)(-0.62,19.38)
\psline{->}(4.38,12.5)
(8.12,16.25)
(8.12,17.5)(8.12,19.38)
\rput{0}(23.12,4.38){\psellipse[fillstyle=solid](0,0)(1.56,-1.56)}
\pscustom[]{\psline{-}(23.12,0)(23.12,-0.62)
\psbezier(23.12,-0.62)(23.12,-0.62)(23.12,-0.62)
\psline{->}(23.12,-0.62)(23.12,3.12)
}
\psline{->}(23.12,3.75)
(20,6.88)
(18.75,8.12)(18.75,11.25)
\psline{->}(23.75,4.38)
(27.5,8.12)
(27.5,9.38)(27.5,11.25)
\pscustom[]{\psline{-}(18.75,15)(18.75,15.62)
\psbezier(18.75,15.62)(18.75,15.62)(18.75,15.62)
\psline{->}(18.75,15.62)(18.75,19.38)
}
\pscustom[]{\psbezier{-}(27.5,15)(27.5,15)(27.5,15)(27.5,15)
\psbezier(27.5,15)(27.5,15)(27.5,15)
\psline{->}(27.5,15)(27.5,19.38)
}
\pscustom[]{\psbezier{-}(51.25,0)(51.25,0)(51.25,0)(51.25,0)
\psbezier(51.25,0)(51.25,0)(51.25,0)
\psline{->}(51.25,0)(51.25,5.62)
}
\rput{0}(51.25,14.38){\psellipse[fillstyle=solid](0,0)(1.56,-1.56)}
\pscustom[]{\psline{-}(51.25,10)(51.25,9.38)
\psbezier(51.25,9.38)(51.25,9.38)(51.25,9.38)
\psline{->}(51.25,9.38)(51.25,13.12)
}
\rput{0}(65,14.38){\psellipse[fillstyle=solid](0,0)(1.56,-1.56)}
\psline{->}(65,10)
(65,0)
(65,9.38)(65,13.12)
\end{pspicture}

\end{center}
If $\CCC$ is the category of relations, then the first equation says that $f$ is a single-valued relation, whereas the second equation says that it is total. Hence the name. 

{\it Notation.} Assuming that $\CCC$ is given with a chosen data service on every objects, we denote by $\CCC^\natural$ the category of functions in $\CCC$. 

Note that the morphisms $\cmn$ and $\cun$ from the data services are functions with respect to the induced data services. They are thus contained in $\CCC^\natural$, and they are natural with respect to the functions. It follows that $\CCC^\natural$ is a cartesian category.
%
%
%
%
%
%
%
%
%
%
%

\section{Monoidal computer}\label{Sec-MonCom}
\subsection{Background and definition}
From \cite{LawvereFW:adjf} to \cite{Lambek-Scott:book}, the structure of \emph{cartesian closed categories} has been the foundation of categorical logic and type theory. It is based on the correspondence:
\bea\label{ccc}
\CCC(X, B^A) & \cong & \CCC(X\times A, B)
\eea
between the functions $X\times A\tto{f(x,a)} B$ on the right hand side, and their abstractions $X\tto{\lambda a. f(x,a)} B^A$ on the left. In a practice-oriented semantics of computation, the $\lambda$-abstraction could be used to represent programming, as an operation mapping the specifications of the computable functions on the right to the programs on the left.  The program evaluation, as an operation mapping the programs to computable functions, would then be represented by the application, i.e. the transition from left to right in \eqref{ccc}. But since the correspondence in \eqref{ccc} is bijective, these transitions don't just evaluate each program into a unique computable function, but also assign a unique program to each computable function. In reality, of course, there are always infinitely many different programs that compute the same function, as soon as the programming language can express enough arithmetic. The \emph{extensional}\/ models of computation, underlying \eqref{ccc}, can be interpreted as viewing the computer as a black box, where only the inputs and the outputs are observable, and any two programs that map the same inputs to the same outputs are indistinguishable. Such view has been not only the tenet of the theory of \emph{denotational}\/ semantics \cite{ScottD:domains,Stoy}, but also the stepping stone into the practice of functional programming \cite{ThompsonS:book}. 

The \emph{intensional}\/ view of computation can also be presented categorically in many ways \cite{eilenberg-elgot,heller-dipaola,Cockett-Hofstra}, as mentioned in the Introduction. One obvious way to allow multiple programs for the same computable function is to relax the bijection in \eqref{ccc} to a surjection. This surjection is the main structural component of monoidal computer. After some fine tuning, it takes the form
\bea\label{moncomp}
\CCC^\natural(X, \PP) & \eepi{\gamma^{AB}_X} & \CCC(X\otimes A, B)
\eea
where the program enumerations $X\tto {F_x} \PP$ on the left are mapped to the computations $X\otimes A \tto{\{F_x\}(a)} B$ on the right. Here the Kleene bracket $\{-\}$ executes the program $F_x$ and yields the computable function $f = \{F_x\}:A\to B$, which can be applied to all data $a$ of type $A$. To understand the step from \eqref{ccc} to \eqref{moncomp}, consider the category  $\Pfn$ of partial computable functions between the finite powers of the set of natural numbers $\NNn$, with $1=\NNn^0$. Since $\Pfn(X,1) = \WP X$, there is no terminal object, thus no cartesian structure. That is why cartesian products $\times$ from \eqref{ccc} are relaxed to the tensor products $\otimes$ in \eqref{moncomp}. To be able to copy and to delete the data, we must specify the data services explicitly. Note that this does not just recover the cartesian structure without the naturality requirement, as there are generally many nonstandard data services, already in the categories as standard as $\Rel$ \cite{PavlovicD:QI09}. While the partiality can be modeled within the cartesian structure \cite{heller-dipaola,Cockett-Hofstra}, going beyond the standard model of computation, and modeling the randomized and quantum computers, does seem to genuinely require nonstandard data services \cite{PavlovicD:CQstruct,PavlovicD:QPL09,PavlovicD:Qabs12}. On the other hand, the program enumerations and transformations on the left hand side of \eqref{moncomp} are required to be total and single valued; hence the restriction to $\CCC^\natural$ (denoted $\CCC_c$ in \cite{PavlovicD:CQstruct}). Anticipating that the computations will be typed, but that all programs as data will be of the same type, we also replace the exponent $B^A$ in \eqref{ccc} by the type $\PP$ of programs in \eqref{moncomp}. These intuitions motivate the formal definition of monoidal computer.

\be{defn}\label{def:moncomp}
A \emph{monoidal computer} is a strict symmetric monoidal category $\CCC$ with the following structure for all objects $A$ and $B$ in $\CCC$
\begin{enumerate}[(I)]
\item a data service $A A \overset{\mnd}{\underset{\cmn}{\rightleftarrows}} A \stackrel\cun\rightarrow I$
\item\label{itemgamma} a distinguished \emph{object of programs}\/ $\PP$ and a family $\CCC^\natural(X, \PP) \eepi{\gamma^{AB}_X} \CCC(X A, B)$  of surjections natural in $X$.
\end{enumerate}
\ee{defn}

\subsection{Universal and partial evaluators}\label{sec-uni-part}
The families of surjections \eqref{moncomp} turn out to be just a categorical version of the  \emph{acceptable enumerations}\/ of computable functions \cite{RogersH:book}, where the enumeration indices represent the programs.

\be{prop}\label{prop-us}
Let $\CCC$ be a symmetric monoidal category with data services. Then specifying the surjections ${\gamma^{AB}_X}:\CCC^\natural(X, \PP) \epi  \CCC(X A, B)$ that make $\CCC$ into a monoidal computer is equivalent to giving for all objects $A$ and $B$ in $\CCC$

\begin{enumerate}[(a)]

\item \label{itemuniv} \emph{universal evaluators} 
$\UK^{AB} \in \CCC(\PP  A,B)$ such that for every $f\in \CCC(A, B)$ there is $F\in \CCC^\natural (\PP)$ (which we call a \emph{program} for $f$) such that

\bea\label{eq-universal-evaluator}
 \def\JPicScale{.55}\newcommand{\aah}{A}\renewcommand{\dh}{B}\newcommand{\ahh}{\scriptstyle f} 
\ifx\JPicScale\undefined\def\JPicScale{1}\fi
\psset{unit=\JPicScale mm}
\psset{linewidth=0.3,dotsep=1,hatchwidth=0.3,hatchsep=1.5,shadowsize=1,dimen=middle}
\psset{dotsize=0.7 2.5,dotscale=1 1,fillcolor=black}
\psset{arrowsize=1 2,arrowlength=1,arrowinset=0.25,tbarsize=0.7 5,bracketlength=0.15,rbracketlength=0.15}
\begin{pspicture}(0,0)(13.75,20)
\rput(5,20){$\dh$}
\psline{->}(5,-8.25)(5,-1.25)
\psline{->}(5,8.75)(5,15.5)
\rput(5,3.75){$\ahh$}
\rput(5,-13.12){$\aah$}
\psline(-3.76,4.99)
(-3.75,-1.25)
(13.75,-1.26)
(13.75,8.74)
(-3.76,8.74)
(13.75,8.74)
(-3.76,8.74)(-3.76,4.99)
\end{pspicture}
 & = & \def\JPicScale{.55}\newcommand{\aah}{\scriptscriptstyle F}\renewcommand{\dh}{B}\newcommand{\ahh}{\scriptstyle \UK^{AB}}\newcommand{\bhh}{A} 
\ifx\JPicScale\undefined\def\JPicScale{1}\fi
\psset{unit=\JPicScale mm}
\psset{linewidth=0.3,dotsep=1,hatchwidth=0.3,hatchsep=1.5,shadowsize=1,dimen=middle}
\psset{dotsize=0.7 2.5,dotscale=1 1,fillcolor=black}
\psset{arrowsize=1 2,arrowlength=1,arrowinset=0.25,tbarsize=0.7 5,bracketlength=0.15,rbracketlength=0.15}
\begin{pspicture}(0,0)(23.12,20)
\rput(16.88,20){$\dh$}
\psline{->}(6.88,-0.62)(6.88,3.12)
\psline{->}(16.88,8.12)(16.88,15.62)
\rput(16.88,3.12){$\ahh$}
\rput(5,-2.5){$\aah$}
\psline{->}(17.5,-8.11)(17.5,-1.88)
\rput(17.5,-12.5){$\bhh$}
\pspolygon(1.88,4.36)
(11.86,-5.62)
(1.88,-5.62)(1.88,4.36)
\psline(1.88,8.12)
(11.88,-1.87)
(23.12,-1.88)
(23.12,8.12)
(5.62,8.12)
(23.12,8.12)
(5.62,8.12)(1.88,8.12)
\end{pspicture}

\eea
\vspace{1\baselineskip}

\item \label{itempart}
\emph{partial evaluators} 
$\SK^{(AB)C} \in \CCC^\natural(\PP  A,\PP)$ such that 

\bea\label{eq-partial-evaluator}
\def\JPicScale{.65}\renewcommand{\dh}{\PP}\newcommand{\ahh}{\scriptstyle \UK^{(AB)C}}\newcommand{\bhh}{A}\newcommand{\chh}{B}\newcommand{\Lhh}{C} 
\ifx\JPicScale\undefined\def\JPicScale{1}\fi
\psset{unit=\JPicScale mm}
\psset{linewidth=0.3,dotsep=1,hatchwidth=0.3,hatchsep=1.5,shadowsize=1,dimen=middle}
\psset{dotsize=0.7 2.5,dotscale=1 1,fillcolor=black}
\psset{arrowsize=1 2,arrowlength=1,arrowinset=0.25,tbarsize=0.7 5,bracketlength=0.15,rbracketlength=0.15}
\begin{pspicture}(0,0)(25.62,14.38)
\rput(4.38,-11.25){$\dh$}
\psline{->}(3.75,-7.5)(3.75,1.88)
\psline{->}(14.38,-7.5)(14.38,-3.12)
\rput(14.38,-11.25){$\bhh$}
\rput(16.25,1.88){$\ahh$}
\psline{->}(21.25,-7.5)(21.25,-3.12)
\rput(21.25,-11.25){$\chh$}
\psline{->}(15.62,6.24)(15.62,11.25)
\rput(15.62,14.38){$\Lhh$}
\psline(-0.62,6.25)
(8.75,-3.12)
(25.62,-3.12)
(25.62,6.25)
(5.62,6.26)
(3.12,6.26)
(1.25,6.25)(-0.62,6.25)
\end{pspicture}
 & = & \def\JPicScale{.65}\renewcommand{\dh}{\Nn}
\newcommand{\ahh}{\scriptstyle \UK^{BC}}\newcommand{\shh}{\scriptstyle \SK^{(AB)C}}\newcommand{\bhh}{A}\newcommand{\chh}{B}\newcommand{\Lhh}{C} 
\ifx\JPicScale\undefined\def\JPicScale{1}\fi
\psset{unit=\JPicScale mm}
\psset{linewidth=0.3,dotsep=1,hatchwidth=0.3,hatchsep=1.5,shadowsize=1,dimen=middle}
\psset{dotsize=0.7 2.5,dotscale=1 1,fillcolor=black}
\psset{arrowsize=1 2,arrowlength=1,arrowinset=0.25,tbarsize=0.7 5,bracketlength=0.15,rbracketlength=0.15}
\begin{pspicture}(0,0)(30.01,13.12)
\psline{->}(2.5,-11.88)(2.5,-3.1)
\psline{->}(18.13,-2.5)(18.13,1.25)
\psline{->}(15,-11.88)(15,-7.5)
\rput(15,-15.62){$\bhh$}
\rput(24.37,1.25){$\ahh$}
\rput(11.25,-3.12){$\shh$}
\psline{->}(26.87,-12.5)(26.87,-3.75)
\rput(26.87,-15.62){$\chh$}
\rput(2.5,-15.62){$\dh$}
\psline{->}(26.87,5.62)(26.87,10.62)
\rput(26.87,13.12){$\Lhh$}
\psline(13.76,5.62)
(23.13,-3.75)
(30.01,-3.75)
(30.01,5.62)
(26.87,5.62)
(16.87,5.62)
(15,5.62)(13.76,5.62)
\pscustom[]{\psline(-2.5,1.88)(6.88,-7.5)
\psline(6.88,-7.5)(23.13,-7.5)
\psline(23.13,-7.5)(16.86,-1.25)
\psline(16.86,-1.25)(13.74,1.88)
\psline(13.74,1.88)(-2.5,1.88)
\psline(-2.5,1.88)(-1.25,1.88)
\psbezier(-1.25,1.88)(-1.25,1.88)(-1.25,1.88)
}
\end{pspicture}

\eea
\vspace{1\baselineskip}
\end{enumerate}
\ee{prop}


\be{prop}\label{prop-retracts}
Every type $A$ in a monoidal computer is a retract of $\PP$: there are computations $\iota^A\in \CCC(A,\PP)$ and $\upsilon^A\in \CCC(\PP,A)$, such that the composite $A\tto{\iota^A}\PP \eepi{\upsilon^A} A$ is the identity. These retractions are isomorphisms if and only if the family of surjections is natural in $A$ or in $B$. Monoidal computer then provides a model of untyped $\lambda$-calculus.
\ee{prop}

{\it Remark.} In \cite{PavlovicD:IC12} we only considered the \emph{basic}\/ monoidal computer, where all types were the powers of $\PP$. In the standard model, the programs are encoded as natural numbers, and all data are the tuples of natural numbers. Prop.~\ref{prop-retracts} implies that all types must also be recursively enumerable in the internal sense of $\CCC$. In particular, by extending the $\lambda$-calculus constructions used in \cite{PavlovicD:IC12}, we can extract from $\PP$ the convenient types of natural numbers, truth values, etc. Here we only need the booleans. Deriving the boolean operations from the structure of the monoidal computer is an instructive exercise.

\be{defn} \label{def-pred} A \emph{predicate}\/ in a monoidal computer $\CCC$ is a computation $\alpha \in \CCC(A,\Bits)$, where the type $\Bits$ of \emph{booleans} is a type where $\CCC^\natural (\Bits) \cong \{0,1\}$. 
\ee{defn}

\subsection{Examples of monoidal computer}
The standard model of monoidal computer is the category of partial computable functions over the tuples of natural numbers. The programs are also encoded as natural numbers, and thus $\PP = \NNn$. The objects can be just the powers of $\NNn$, but also all of their recursively enumerable subsets. The universal evaluators can be the computations implemented by a fixed family of universal Turing Machines. The partial evaluators are the total recursive functions constructed in Kleene's \emph{s-m-n}-Lemma \cite{KleeneSC1936}. Extending this model to recursive relations and nondeterministic Turing machines leads to minor changes of the evaluation structure, but introduces many nonstandard data services \cite{PavlovicD:QI09}, which can be used to encode nonstandard algorithms \cite{PavlovicD:Qabs12}. A \emph{quantum}\/ monoidal computer can be defined within the category of complex Hilbert spaces, with all linear maps as morphisms.  The data services are provided by Frobenius algebras \cite{PavlovicD:QMWS}. The category of functions with respect to these data services is equivalent with the category of sets and functions \cite{PavlovicD:CQstruct}, so the encoding of programs will be classical, and the same as in the standard monoidal computer. The universal and the partial evaluators can be defined as in \cite{Bernstein-Vazirani}. It is important to note, however, that the program evaluations $\gamma^{AB}$ are not surjective in a set-theoretic sense, but dense for the topological sense spelled out in \cite{Bernstein-Vazirani}. Lastly, let us mention that any  \emph{reflexive domain}\/ \cite{MisloveM:compendium} gives rise to an \emph{extensional}\/ monoidal computer, which embodies both \eqref{ccc} and \eqref{moncomp}. For more detail see \cite[Sec.~4.1]{PavlovicD:IC12}.

\section{Graded monoidal computer}\label{Sec-Grad}
\subsection{Graded categories}
While categories are very convenient for denotational semantics, capturing computation as a process requires some additional structure \cite{AbramskyS:retracing,PavlovicD:CTCS97,CatSOS}. Capturing the complexity of computations requires a quantifiable view of that process, to allow us to count the steps, measure the memory, etc. If the morphisms of a monoidal computer represent computations, then we need to introduce some structure over these morphisms to express how much of a computational resource each of them uses. One idea is to consider the subsets $\CCC_n(A,B)\subseteq \CCC(A,B)$ that consist of those computations that use at most $n$ units of a given resource: e.g., at most $n$ steps in time, or $n$ cells of space. Since the composite programs $\comp p q$ and $p\otimes q$ may need up to $m\opls n$ units of the resource, if $p$ needs $m$ units, and $q$ needs $n$, then the scale in which the resource will be measured must carry at least the structure of an additive monoid. 
 
\subsubsection*{Grading monoids} 
Let $(\MMMM  , \opls, 0, \infty)$ be a commutative monoid with the absorptive element  $\infty$, i.e. such that $\infty \opls  m = \infty$ for all $m\in \MMMM  $. The relation
\bea\label{pord}
m\leq n & \iff & \exists \ell.\  \ell \opls  m = n 
\eea
is obviously transitive and reflexive, i.e. a \emph{preorder}. The equivalence classes with respect to the relation $m\sim n \ \iff\ m\leq n \wedge m\geq n$ are also the factors of the subgroup $\GGG = \{m\in \MMMM  \ |\ \exists n.\ m\opls n = 0\}$. For simplicity, we assume $\GGG = \{0\}$, i.e. begin by factoring $\MMMM  $ modulo $\sim$. This not only makes $\leq$ into a partial order, but when $\MMMM  $ is finitely generated, then $(\MMMM  ,\leq)$ is also a well founded lattice.  For more see \cite{SchutzenbergerM:65}. 


\subsubsection*{Examples of grading monoids} The additive monoid of natural numbers completed at the top $\NNnn = \NNn\cup \{\infty \}$ is the free grading monoid over one generator. On the other hand, the monoid of multisets of well-formed expressions in any given language, extended by $\infty$ again, can also be used as a grading monoid. In-between these extremes are the grading monoids normally used in complexity theory: the quotients $\NNnn^\NNn/ \stackrel + \equiv$ and $\NNnn^\NNn/ \stackrel \OOO \equiv$ of the monoid $\NNnn^\NNn$ of functions from $\NNn$ to $\NNnn$, identified modulo the equivalence relations
\bear
f \stackrel + \equiv g & \iff & f\stackrel + \leq g \wedge f \stackrel + \geq g\\
f \stackrel \OOO \equiv g & \iff & f\stackrel \OOO \leq g \wedge f \stackrel \OOO \geq g
\eear
where
\bear
f \stackrel + \leq g & \iff & \exists c\ \forall x.\  f(x) \leq c+ g(x)\\
f \stackrel \OOO \leq g & \iff & \exists cd\ \forall x\geq d.\  f(x) \leq cg(x)
\eear

\be{defn}\label{def-gradcat}
A category $\CCC$ is \emph{$\MMMM  $-graded}\/ if every hom-set $\CCC(A,B)$ comes with 
\begin{enumerate}[(i)]
\item  \emph{grading} $\CCC(A,B) \tto{\grad{-}} \MMMM  $, which induces the graded hom-sets
\bea\label{CCCn}
\CCC_n(A,B)\ = \ \{f\in \CCC(A,B)\ |\  \grad f \leq n\}
\eea
the elements of which we write as $f_n$;
\item \emph{restriction} $\restriction_n\  : \ \CCC (A,B) \to \CCC_n(A,B)$ for every $n\in \NNnn$ where for all $f_m\in \CCC_m(A,B)$ holds
\bea
(f_m) \restr_n & = & f_{m\wedge n}
\eea
\end{enumerate}
We require that identity morphisms are of grade 0, and
\bea  \label{comp-dist}
\grad{\comp f g} & \leq & \grad f \opls  \grad g
\eea
which makes the following diagram commute
\beq
\begin{tikzcd}[ampersand replacement = \&]
\CCC_\ell (A,B) \times \CCC_n(B,C) \arrow[rightarrow]{r}{\compp} 
\arrow[rightarrow]{d}[swap]{\id \times \restriction_m}
\& \CCC_{\ell\opls n}(A,C) 
\arrow[rightarrow]{d}{\restriction_{\ell \opls  m}} 
\\  
\CCC_\ell (A,B) \times \CCC_m(B,C) \arrow[rightarrow]{r}{\compp}
\& \CCC_{\ell\opls m}(A,C)
\end{tikzcd}
\eeq
%
%
%

An \emph{$\MMMM$-graded monoidal category} is a monoidal category $\CCC$, which is $\MMMM$-graded in the above sense, with all monoidal isomorphisms (such as $A\otimes B \tto \varsigma B\otimes A$) of grade 0,  and moreover
\bea\label{otimes-dist}
\grad{f \otimes t} & \leq&  \grad f \opls  \grad t
\eea
making the following diagram commute
%
%
\beq
\begin{tikzcd}[ampersand replacement = \&,column sep=small]
\CCC_\ell (A,B) \times \CCC_n(P,Q) \arrow[rightarrow]{r}{\otimes} 
\arrow[rightarrow]{d}[swap]{\restriction_k \times \restriction_m}
\& \CCC_{\ell\opls n}(AP, BQ) 
\arrow[rightarrow]{d}{\restriction_{k \opls  m}} 
\\  
\CCC_k (A,B) \times \CCC_m(P,Q) \arrow[rightarrow]{r}{\otimes}
\& \CCC_{k\opls m}(AP,BQ)
\end{tikzcd}
\eeq
\ee{defn}

{\it Remark.} Note that we do not impose any requirements on the grading map $\CCC(A,B) \tto{\grad{-}} \MMMM  $ that would allow lifting the lattice structure from $\MMMM  $ to $\CCC(A,B)$, and in general $\bigvee \{n\lt \infty\}=\infty$ does \emph{not}\/ imply that $\bigcup_{n\lt \infty} \CCC_n(A,B)$ covers $\CCC_\infty(A,B) =\CCC(A,B)$. It usually does not.

\be{defn}\label{def-halt}
We say that a computation $I\tto a A$ \emph{halts} if there is $m\lt \infty$ such that $a = a\restr_m$. More generally, we say that $A\tto f B $ \emph{halts}\/ if the composite $I\tto a A\tto f B$ halts whenever $a$ halts. The subcategory of \emph{total functions that halt}\/ is denoted $\CCC_\natural$, and thus 
\bear
\CCC_\natural (A,B) & = & \bigcup_{m\lt \infty} \CCC_m^\natural (A,B)
\eear
\ee{defn}

\be{defn}\label{def-decidable}
We say that a predicate $A \tto \alpha \Bits$ is \emph{decidable}\/ if it always halts and moreover outputs a single value, i.e. $\alpha\in \CCC_\natural$.
\ee{defn}


\subsection{Definition and characterization}
\be{defn}\label{def-gmc}
An $\MMMM$-graded monoidal category $\CCC$ with data services in $\CCC^\natural_0$ is an \emph{$\MMMM$-graded monoidal computer}\/ if it carries 
the following structure for all $A, B\in \CCC$:
\begin{enumerate}[(I)]
\item a family $\CCC^\natural_0(X,\PP)\tto{\gamma^{AB}_X} \CCC(XA,B)$  of surjections natural in $X$,

\item\label{def-grad-II} families $\sigma_X^A$, $\tau_X^A$ and $\vartheta_X^A$, all natural in $X$, such that the following diagrams commute
\bear
\begin{tikzcd}[ampersand replacement = \&,row sep=tiny
]
\CCC_0^\natural (X,\PP)  \arrow[twoheadrightarrow]{rd}[near start]{\gamma_X^{(AB)C}} 
\arrow[rightarrow]{dd}[swap]{\sigma_X^{A}} \\
\& 
\CCC(X AB, C)
\\
\CCC_0^\natural (X A, \PP)
\arrow[twoheadrightarrow,swap]{ur}[near start]{\gamma_{XA}^{BC}} 
\\
%
%
\\
\CCC_0^\natural (X,\PP)  \arrow[twoheadrightarrow,near start]{rd}{\gamma_X^{AB}} 
\arrow[rightarrow]{dd}[swap]{\tau_X^{A}} 
\\
\& 
\CCC(X A, B)
\arrow[twoheadrightarrow]{dd}{\restriction_n}
\\
\CCC(X A, \PP)
\arrow[twoheadrightarrow,swap,near start]{ur}{\vartheta_{XA}^{B}} 
\arrow[twoheadrightarrow]{dd}[swap]{\restriction_n}  \\
\& 
\CCC_n(X A, B)
\\
\CCC_n (X A, \PP)
\arrow[twoheadrightarrow,swap,near start]{ur}{\vartheta_{XAn}^{B}} 
\end{tikzcd}
\eear
\end{enumerate}
\ee{defn}

\be{prop}\label{prop-ustw}
Let $\CCC$ be an $\MMMM$-graded monoidal category with data services. Then specifying the structure of an $\MMMM$-graded monoidal computer as in Def.~\ref{def-gmc}(I-II) is equivalent to giving for all $A,B\in \CCC$ and all $n\in \MMMM  $
 
\begin{enumerate}[(a)]

\item \label{item:u} \emph{graded universal evaluators} 
$\UK_n^{AB} \in \CCC_n(\PP A,B)$ such that for every $f_n\in \CCC_n(A, B)$ there is $F\in \CCC_0^\natural (X,\PP)$ such that
\bea\label{eq-grad-universal-evaluator}
 \def\JPicScale{.6}\newcommand{\aah}{A}\renewcommand{\dh}{B}\newcommand{\ahh}{\scriptstyle f_n} 
\ifx\JPicScale\undefined\def\JPicScale{1}\fi
\psset{unit=\JPicScale mm}
\psset{linewidth=0.3,dotsep=1,hatchwidth=0.3,hatchsep=1.5,shadowsize=1,dimen=middle}
\psset{dotsize=0.7 2.5,dotscale=1 1,fillcolor=black}
\psset{arrowsize=1 2,arrowlength=1,arrowinset=0.25,tbarsize=0.7 5,bracketlength=0.15,rbracketlength=0.15}
\begin{pspicture}(0,0)(13.75,20.62)
\rput(5,20.62){$\dh$}
\psline{->}(5,-8.25)(5,-1.25)
\psline{->}(5,10)(5,16.75)
\rput(5,4.38){$\ahh$}
\rput(5.62,-12.5){$\aah$}
\psline(-3.75,5.01)
(-3.75,-1.25)
(13.75,-1.25)
(13.74,10.01)
(-0.63,10.01)
(-0.63,6.88)
(-3.76,6.88)(-3.75,5.01)
\end{pspicture}
 & = & \def\JPicScale{.6}\newcommand{\aah}{\scriptscriptstyle F}\renewcommand{\dh}{B}\newcommand{\ahh}{\scriptstyle \UK_n^{AB}}\newcommand{\bhh}{A} 
\ifx\JPicScale\undefined\def\JPicScale{1}\fi
\psset{unit=\JPicScale mm}
\psset{linewidth=0.3,dotsep=1,hatchwidth=0.3,hatchsep=1.5,shadowsize=1,dimen=middle}
\psset{dotsize=0.7 2.5,dotscale=1 1,fillcolor=black}
\psset{arrowsize=1 2,arrowlength=1,arrowinset=0.25,tbarsize=0.7 5,bracketlength=0.15,rbracketlength=0.15}
\begin{pspicture}(0,0)(20.63,21.25)
\rput(14.38,21.25){$\dh$}
\psline{->}(5,-1.25)(5,2.5)
\psline{->}(14.38,9.38)(14.38,16.88)
\rput(14.38,3.12){$\ahh$}
\rput(3.75,-3.12){$\aah$}
\psline{->}(15,-8.11)(15,-1.88)
\rput(15.62,-11.88){$\bhh$}
\pspolygon(1.25,2.5)
(9.36,-5.62)
(1.25,-5.62)(1.25,2.5)
\psline(1.25,6.25)
(9.38,-1.88)
(20.63,-1.88)
(20.62,9.38)
(6.25,9.38)
(6.25,6.25)
(3.12,6.25)(1.25,6.25)
\end{pspicture}

\eea

\bigskip
\bigskip
\item 
\emph{partial evaluators} \label{item:s}
$\SK^{A} \in \CCC^\natural_0(\PP A,\PP)$ such that
\bea\label{eq-grad-partial-evaluator}
\def\JPicScale{.65}\renewcommand{\dh}{\PP}\newcommand{\ahh}{\scriptstyle\UK_{n}^{(AB)C}}\newcommand{\bhh}{A}\newcommand{\chh}{B}\newcommand{\Lhh}{C} 
\ifx\JPicScale\undefined\def\JPicScale{1}\fi
\psset{unit=\JPicScale mm}
\psset{linewidth=0.3,dotsep=1,hatchwidth=0.3,hatchsep=1.5,shadowsize=1,dimen=middle}
\psset{dotsize=0.7 2.5,dotscale=1 1,fillcolor=black}
\psset{arrowsize=1 2,arrowlength=1,arrowinset=0.25,tbarsize=0.7 5,bracketlength=0.15,rbracketlength=0.15}
\begin{pspicture}(0,0)(28.75,20)
\rput(7.5,-10){$\dh$}
\psline{->}(7.5,-6.88)(7.5,1.87)
\psline{->}(16.25,-6.88)(16.25,-2.5)
\rput(16.25,-10){$\bhh$}
\rput(18.13,2.5){$\ahh$}
\psline{->}(23.75,-6.88)(23.75,-2.49)
\rput(23.75,-10){$\chh$}
\psline{->}(23.75,11.25)(23.75,16.88)
\rput(23.75,20){$\Lhh$}
\psline(3.12,6.25)
(11.87,-2.5)
(28.75,-2.5)
(28.75,11.25)
(18.75,11.25)
(18.75,6.25)
(5.63,6.25)(3.12,6.25)
\end{pspicture}

 & = & \def\JPicScale{.5}\renewcommand{\dh}{\Nn}
\newcommand{\ahh}{\scriptstyle\UK_{n}^{BC}}\newcommand{\shh}{\scriptstyle \SK^{A}}\newcommand{\bhh}{A}\newcommand{\chh}{B}\newcommand{\Lhh}{C} 
\ifx\JPicScale\undefined\def\JPicScale{1}\fi
\psset{unit=\JPicScale mm}
\psset{linewidth=0.3,dotsep=1,hatchwidth=0.3,hatchsep=1.5,shadowsize=1,dimen=middle}
\psset{dotsize=0.7 2.5,dotscale=1 1,fillcolor=black}
\psset{arrowsize=1 2,arrowlength=1,arrowinset=0.25,tbarsize=0.7 5,bracketlength=0.15,rbracketlength=0.15}
\begin{pspicture}(0,0)(31.25,18.75)
\psline{->}(8.12,-13.12)(8.12,-3.12)
\psline{->}(16.88,-2.5)(16.88,1.25)
\psline{->}(16.88,-11.88)(16.88,-7.64)
\rput(16.88,-16.25){$\bhh$}
\rput(25.62,1.88){$\ahh$}
\rput(12.5,-2.5){$\shh$}
\psline{->}(26.88,-11.88)(26.88,-3.75)
\rput(29.38,-16.25){$\chh$}
\rput(8.12,-16.25){$\dh$}
\psline{->}(26.25,10)(26.25,15.62)
\rput(26.25,18.75){$\Lhh$}
\psline(6.25,-1.25)
(12.5,-7.5)
(21.88,-7.5)
(15.62,-1.25)
(12.5,1.88)
(3.12,1.88)
(6.25,-1.25)(10,-5)
\psline(13.11,4.99)
(21.86,-3.76)
(31.25,-3.75)
(31.25,9.99)
(21.24,10)
(21.24,5)
(15.62,5)(13.11,4.99)
\end{pspicture}
 
\eea
\vspace{1\baselineskip}

\item  \emph{trace evaluators} \label{item:tw}
$\TK_n \in \CCC_n(\PP,\PP)$ and\\ 
\emph{output extractors} 
$\VK^B \in \CCC^\natural_0(\PP,B)$ such that 
\nopagebreak
\vspace{.5\baselineskip}
\bea\label{eq-total-evaluator}
\def\JPicScale{.7}\renewcommand{\dh}{\PP}\newcommand{\ahh}{\scriptstyle\UK_n^{IB}}\newcommand{\chh}{\PP}\newcommand{\Lhh}{B} 
\ifx\JPicScale\undefined\def\JPicScale{1}\fi
\psset{unit=\JPicScale mm}
\psset{linewidth=0.3,dotsep=1,hatchwidth=0.3,hatchsep=1.5,shadowsize=1,dimen=middle}
\psset{dotsize=0.7 2.5,dotscale=1 1,fillcolor=black}
\psset{arrowsize=1 2,arrowlength=1,arrowinset=0.25,tbarsize=0.7 5,bracketlength=0.15,rbracketlength=0.15}
\begin{pspicture}(0,0)(10.62,15.62)
\rput(5.62,-15){$\dh$}
\rput(6.88,0.62){$\ahh$}
\psline{->}(5.62,-10.62)(5.62,-5.62)
\psline{->}(6.25,6.87)(6.25,11.88)
\rput(6.25,15.62){$\Lhh$}
\pscustom[]{\psline(1.88,-1.88)(10.62,-10.62)
\psbezier(10.62,-10.62)(10.62,-10.62)(10.62,-10.62)
\psline(10.62,-10.62)(10.62,6.88)
\psline(10.62,6.88)(1.88,6.88)
\psline(1.88,6.88)(1.88,-1.88)
\psbezier(1.88,-1.88)(1.88,-1.88)(1.88,-1.88)
\psline(1.88,-1.88)(1.88,5.62)
}
\end{pspicture}
 & = & \def\JPicScale{.7}\renewcommand{\dh}{\Nn}
\newcommand{\ahh}{\scriptscriptstyle \VK^B}\newcommand{\shh}{\scriptstyle \TK_n}\newcommand{\Lhh}{B} 
\ifx\JPicScale\undefined\def\JPicScale{1}\fi
\psset{unit=\JPicScale mm}
\psset{linewidth=0.3,dotsep=1,hatchwidth=0.3,hatchsep=1.5,shadowsize=1,dimen=middle}
\psset{dotsize=0.7 2.5,dotscale=1 1,fillcolor=black}
\psset{arrowsize=1 2,arrowlength=1,arrowinset=0.25,tbarsize=0.7 5,bracketlength=0.15,rbracketlength=0.15}
\begin{pspicture}(0,0)(15,16.25)
\rput(9.38,-18.75){$\dh$}
\rput(12,6.38){$\ahh$}
\psline{->}(10,-15)(10,-9.37)
\psline{->}(10.62,8.76)(10.62,13.12)
\rput(10.62,16.25){$\Lhh$}
\pscustom[]{\psline(5.01,-4.38)(15,-14.36)
\psbezier(15,-14.36)(15,-14.36)(15,-14.36)
\psline(15,-14.36)(15,-5.62)
\psline(15,-5.62)(5.01,4.38)
\psline(5.01,4.38)(5.01,-4.38)
\psbezier(5.01,-4.38)(5.01,-4.38)(5.01,-4.38)
\psline(5.01,-4.38)(5.01,2.5)
}
\rput(9.38,-5){$\shh$}
\psline(5,8.76)
(15,8.76)
(15,-1.24)
(13.75,-0.01)(5,8.76)
\psline{->}(10,-0.62)(10,3.75)
\end{pspicture}
 
\eea
\vspace{1\baselineskip}
\end{enumerate}
\ee{prop}

\subsection{Examples of graded monoidal computer}
First the bad news. Since graded monoidal computer extends the structure of monoidal computer, it seems natural that the standard model of monoidal computer, with a suitable grading, should provide a model of graded monoidal computer. However, the standard monoidal computer cannot be extended to a graded monoidal computer! The reason lies in Blum's Speedup Theorem \cite{BlumM:axioms,BlumM:speedup}. Blum's constructed  a computable function such that for every program that implements it there is another program that computes the same, but faster by an arbitrary recursive factor. The construction applies to an arbitrary Blum measure (cf. Def.~\ref{def-blum}). The message is that complexity is not a property of computable functions, but of programs. Therefore, the category of computable functions, which provides the standard model of monoidal computer, is not a good place to measure complexity. 

The good news is that this is not a bug, but an important feature of  the approach! The concept of graded monoidal computer, as a categorical axiomatization of computational complexity, uncovers the fact that the universe of computation should not be modeled as the category of computable functions, but the category of computations, viewed as the pairs $\left<\mbox{program},\mbox{function that it implements}\right>$. The realisation that this is the right categorical model echoes what Levin and Meyer called the \emph{"Fundamental Theorem of Complexity Theory"}\/\cite{LevinL:fundamental,MeyerA:speedup}, which they formulated as a statement unifying Blum's Speedup and Compression Theorems \cite{BlumM:axioms,BlumM:speedup}. 
The standard model of graded monoidal computer thus presents a computation as a pairs $A\tto{<f,F>} B$, where\footnote{Recall that Kleene's bracket $\{-\}$ is used in classical resursion theory to denote the universal evaluators.} $\{F\} = f$. This is a monoidal computer, since for every computation $A\tto{<f,F>} B$, i.e. for every program $F$, there is a program $\Phi$ such that $\{\Phi\}=F$ and a morphism $1\tto{<F,\Phi>} \PP$, so that we can define $\gamma^{AB}_I(F,\Phi) = \left<\{F\}, \{\Phi\}\right> = <f,F>$. 

To define the grading, consider the set $\TTT$ of the execution traces like \eqref{trace}, and fix an encoding $\TTT \tto{\enco{-}}\NNn$ such that the parallel and the sequential compositions of the traces are associative and unitary. These requirements mean that the program compositions, the data-passing and buffering operations (both modeled by the identity morphisms), and the input-output operations are all assigned grade 0. Define a total recursive function $\NNn\times \NNn\tto \oplus \NNn$ so that $\enco f \oplus \enco g \geq \enco{\comp f g}$ and $\enco f \oplus \enco t \geq \enco{f \otimes t}$. Fix the universal evaluators $\UK$, and define $\UK_n^{AB}(F,a)$ to not just build the trace $\TK(F,a)$, but at the same time computes the code $\enco{\TK(F,a)}\in \NNn$; and that it halts when $\enco{\TK(F,a)}\gt n$. Let $\CCC_n(A,B)$ consist of the computations $<f,F>$ that "clock out" at $n$, i.e. $\UK(F,a) = \UK_n(F,a)$ for all $a$.

While the details of the trace encodings and their use in the evaluations have to be deferred for the full paper, it should be clear that they amount to a routinely, albeit lengthy programming task, that awaits, e.g., the designer of a debugging tool that needs to capture, store, and play the internal execution traces. 

The nondeterministic and the quantum monoidal computers lift in a similar way, but lead to substantially different grading structures. 

\section{Normal form}\label{Sec-Norm}
\be{lemma}
Every graded universal evaluator $\UK_n^{AB}$ in a graded monoidal computer decomposes into the normal form $\SK^A\, ; \TK_n\, ; \VK^B$, or diagrammatically:

\bigskip
\bea\label{eq-u-decomp}
\def\JPicScale{.6}\renewcommand{\dh}{\Nn}
\newcommand{\bhh}{A} \newcommand{\Lhh}{B} \newcommand{\ahh}{\scriptstyle \UK_n^{AB}} 
\ifx\JPicScale\undefined\def\JPicScale{1}\fi
\psset{unit=\JPicScale mm}
\psset{linewidth=0.3,dotsep=1,hatchwidth=0.3,hatchsep=1.5,shadowsize=1,dimen=middle}
\psset{dotsize=0.7 2.5,dotscale=1 1,fillcolor=black}
\psset{arrowsize=1 2,arrowlength=1,arrowinset=0.25,tbarsize=0.7 5,bracketlength=0.15,rbracketlength=0.15}
\begin{pspicture}(0,0)(25.62,23.12)
\rput(19.38,0.62){$\ahh$}
\psline{->}(20.62,13.12)(20.62,18.75)
\rput(20.62,23.12){$\Lhh$}
\pscustom[]{\psline(15,-5.62)(25.62,-5.62)
\psline(25.62,-5.62)(25.62,-5)
\psline(25.62,-5)(25.62,13.12)
\psline(25.62,13.12)(15.62,13.12)
\psline(15.62,13.12)(15.62,4.38)
\psline(15.62,4.38)(15.62,7.5)
\psbezier(15.62,7.5)(15.62,7.5)(15.62,7.5)
}
\psline{->}(8.75,-10)(8.75,-0.62)
\psline{->}(20,-10)(20,-5.76)
\rput(20,-13.75){$\bhh$}
\rput(8.75,-13.75){$\dh$}
\pscustom[]{\psline(3.75,4.38)(13.75,-5.62)
\psline(13.75,-5.62)(25.62,-5.62)
\psline(25.62,-5.62)(20,-5.62)
\psline(20,-5.62)(13.75,-5.62)
\psline(13.75,-5.62)(3.75,4.38)
\psbezier(3.75,4.38)(3.75,4.38)(3.75,4.38)
\psline(3.75,4.38)(15.62,4.38)
}
\end{pspicture}
 & = & \def\JPicScale{.6}\newcommand{\Fhh}{\PP}\newcommand{\ahh}{\scriptscriptstyle\VK^{B}}\newcommand{\bhh}{A}\newcommand{\Lhh}{B}\newcommand{\shh}{\scriptstyle\SK^{A}} \newcommand{\thh}{\scriptstyle \TK_n}
\ifx\JPicScale\undefined\def\JPicScale{1}\fi
\psset{unit=\JPicScale mm}
\psset{linewidth=0.3,dotsep=1,hatchwidth=0.3,hatchsep=1.5,shadowsize=1,dimen=middle}
\psset{dotsize=0.7 2.5,dotscale=1 1,fillcolor=black}
\psset{arrowsize=1 2,arrowlength=1,arrowinset=0.25,tbarsize=0.7 5,bracketlength=0.15,rbracketlength=0.15}
\begin{pspicture}(0,0)(23.12,25)
\rput(20,16.25){$\ahh$}
\psline{->}(18.12,-4.38)(18.13,-0.63)
\psline{->}(18.75,18.12)(18.75,22.5)
\rput(18.75,25){$\Lhh$}
\pscustom[]{\psline(13.75,3.75)(22.5,-5.01)
\psbezier(22.5,-5.01)(22.5,-5.01)(22.5,-5.01)
\psline(22.5,-5.01)(22.5,5.62)
\psline(22.5,5.62)(13.75,14.37)
\psline(13.75,14.37)(13.75,3.75)
\psbezier(13.75,3.75)(13.75,3.75)(13.75,3.75)
\psline(13.75,3.75)(13.75,11.25)
}
\psline(13.75,18.12)
(23.12,18.12)
(23.12,8.75)
(22.5,9.36)(13.75,18.12)
\psline{->}(18.75,9.38)(18.75,13.12)
\psline{->}(6.25,-13.12)(6.25,-4.38)
\psline{->}(16.25,-12.98)(16.25,-8.76)
\rput(16.25,-16.88){$\bhh$}
\rput(11.25,-3.12){$\shh$}
\pscustom[]{\psline(1.88,0)(10.62,-8.76)
\psline(10.62,-8.76)(22.5,-8.76)
\psline(22.5,-8.76)(13.75,0)
\psbezier(13.75,0)(13.75,-0)(13.75,-0)
\psline(13.75,-0)(1.88,-0)
\psbezier(1.88,-0)(1.88,0)(1.88,0)
\psbezier(1.88,0)(1.88,0)(1.88,0)
}
\rput(18.12,4.38){$\thh$}
\rput(6.25,-16.88){$\Fhh$}
\end{pspicture}

\eea
\vspace{2\baselineskip}
\ee{lemma}

\be{corollary} \label{Corr}Every computation $f_n :A\to B$ in a graded monoidal computer decomposes for each of its programs $F$ into the normal form $F\,  ; \SK^A\, ; \TK_n\, ; \VK^B$

\bea\label{eq-f-decomp}
\def\JPicScale{.6}\renewcommand{\dh}{\Nn}
\newcommand{\bhh}{A} \newcommand{\Lhh}{B} \newcommand{\ahh}{\scriptstyle f_n} 
\ifx\JPicScale\undefined\def\JPicScale{1}\fi
\psset{unit=\JPicScale mm}
\psset{linewidth=0.3,dotsep=1,hatchwidth=0.3,hatchsep=1.5,shadowsize=1,dimen=middle}
\psset{dotsize=0.7 2.5,dotscale=1 1,fillcolor=black}
\psset{arrowsize=1 2,arrowlength=1,arrowinset=0.25,tbarsize=0.7 5,bracketlength=0.15,rbracketlength=0.15}
\begin{pspicture}(0,0)(25.62,23.12)
\rput(18.12,0){$\ahh$}
\psline{->}(20.62,13.12)(20.62,18.75)
\rput(20.62,23.12){$\Lhh$}
\pscustom[]{\psline(15,-5.62)(25.62,-5.62)
\psline(25.62,-5.62)(25.62,-5)
\psline(25.62,-5)(25.62,13.12)
\psline(25.62,13.12)(15.62,13.12)
\psline(15.62,13.12)(15.62,4.38)
\psline(15.62,4.38)(15.62,7.5)
\psbezier(15.62,7.5)(15.62,7.5)(15.62,7.5)
}
\psline{->}(18.75,-11.11)(18.75,-5.62)
\rput(18.75,-15.62){$\bhh$}
\psline(3.75,-5.62)
(13.75,-5.62)
(25.62,-5.62)
(20,-5.62)
(13.75,-5.62)
(3.75,-5.62)
(3.75,4.38)(15.62,4.38)
\end{pspicture}
 & = & \def\JPicScale{.6}\renewcommand{\dh}{\PP}\newcommand{\ahh}{\scriptscriptstyle\VK^{B}}\newcommand{\bhh}{A}\newcommand{\Lhh}{B}\newcommand{\shh}{\scriptstyle\SK^A} \newcommand{\thh}{\scriptstyle \TK_n}\newcommand{\Fhh}{\scriptscriptstyle F}
\ifx\JPicScale\undefined\def\JPicScale{1}\fi
\psset{unit=\JPicScale mm}
\psset{linewidth=0.3,dotsep=1,hatchwidth=0.3,hatchsep=1.5,shadowsize=1,dimen=middle}
\psset{dotsize=0.7 2.5,dotscale=1 1,fillcolor=black}
\psset{arrowsize=1 2,arrowlength=1,arrowinset=0.25,tbarsize=0.7 5,bracketlength=0.15,rbracketlength=0.15}
\begin{pspicture}(0,0)(23.12,25)
\rput(20,16.25){$\ahh$}
\psline{->}(18.12,-4.38)(18.13,-0.63)
\psline{->}(18.75,18.12)(18.75,22.5)
\rput(18.75,25){$\Lhh$}
\pscustom[]{\psline(13.75,3.75)(22.5,-5.01)
\psbezier(22.5,-5.01)(22.5,-5.01)(22.5,-5.01)
\psline(22.5,-5.01)(22.5,5.62)
\psline(22.5,5.62)(13.75,14.37)
\psline(13.75,14.37)(13.75,3.75)
\psbezier(13.75,3.75)(13.75,3.75)(13.75,3.75)
\psline(13.75,3.75)(13.75,11.25)
}
\psline(13.75,18.12)
(23.12,18.12)
(23.12,8.75)
(22.5,9.36)(13.75,18.12)
\psline{->}(18.75,9.38)(18.75,13.12)
\psline{->}(6.25,-8.12)(6.25,-4.38)
\psline{->}(16.25,-12.98)(16.25,-8.76)
\rput(16.25,-16.88){$\bhh$}
\rput(11.25,-3.12){$\shh$}
\pscustom[]{\psline(1.88,0)(10.62,-8.76)
\psline(10.62,-8.76)(22.5,-8.76)
\psline(22.5,-8.76)(13.75,0)
\psbezier(13.75,0)(13.75,-0)(13.75,-0)
\psline(13.75,-0)(1.88,-0)
\psbezier(1.88,-0)(1.88,0)(1.88,0)
\psbezier(1.88,0)(1.88,0)(1.88,0)
}
\rput(18.12,4.38){$\thh$}
\rput(5,-10){$\Fhh$}
\psline(1.88,-3.73)
(1.88,-12.48)
(10.63,-12.48)
(10,-11.88)(1.88,-3.73)
\end{pspicture}

\eea
\ee{corollary}
\vspace{2\baselineskip}

{\it Remark} For $n = \infty$, Corollary~\ref{Corr} is the monoidal version of Kleene's Normal Form Theorem \cite[or any textbook in recursion theory]{KleeneSC1936}. Kleene proved his theorem by specifying a primitive recursive output extractor $\VK^B$, and implementing $\TK_\infty\left( \SK^A(F,a)\right)$ in the form $\mu x. T(F,a,x)$, where $T$ is a primitive recursive predicate that verifies that $x$ is the trace of the evaluation $\{F\}(a)$ of the program $F$ on the input $a$. Corollary~\ref{Corr} just spells out that the categorical axioms of graded monoidal computer suffice for Kleene's decomposition. The point is that this decomposition, in a sense, displays the process of computation, encoded in the execution traces, as the grading of the trace evaluators $\TK_n$. This allows us to measure complexity.

\section{Complexity measures}\label{Sec-CX}
\subsection{Internal grading}
Intuitively, a complexity measure is a computable function $\CX$ that takes a program $F$ and an input $a$, and measures how much of a computational resource is needed to evaluate $F$ on $a$. Since the measurements will be derived from the grades, we now assume that the grading monoid $\MMMM$ is representable in $\CCC$. This means that there is an object $\MMmm$ in $\CCC$ such that each $m\in \MMMM$ is just a basis point $I\tto m \MMmm$, i.e. $\MMMM  =  \CCC_0^\natural(\MMmm)$.
The monoid structure of $\MMMM$ is internalized as the diagram $\MMmm\otimes \MMmm \stackrel{\opls}\rightarrow \MMmm \overset{0}{\underset{\infty}{\leftleftarrows}} I$ in $\CCC^\natural_0$. It is easy to see that the  ordering $\leq$ of $\MMMM$ is then also representable as an internal predicate $\MMmm\otimes \MMmm\tto{\olessthan}\Bits$ in $\CCC^\natural_0$.

We further assume that the grading of the trace evaluators $\TK_n\in \CCC_n(\PP,\PP)$ is internalized in the sense that
there is $\TK\in \CCC(\PP\otimes \MMmm, \PP)$ such that
\bea\label{eq-ttn}
\def\JPicScale{.55}\renewcommand{\dh}{\PP}\renewcommand{\dh}{\Nn}
\newcommand{\ahh}{\scriptstyle n}\newcommand{\shh}{\scriptstyle \TK_n}\newcommand{\Lhh}{\MMmm} 
\ifx\JPicScale\undefined\def\JPicScale{1}\fi
\psset{unit=\JPicScale mm}
\psset{linewidth=0.3,dotsep=1,hatchwidth=0.3,hatchsep=1.5,shadowsize=1,dimen=middle}
\psset{dotsize=0.7 2.5,dotscale=1 1,fillcolor=black}
\psset{arrowsize=1 2,arrowlength=1,arrowinset=0.25,tbarsize=0.7 5,bracketlength=0.15,rbracketlength=0.15}
\begin{pspicture}(0,0)(15,11.88)
\rput(9.38,-13.12){$\dh$}
\psline{->}(10,-9.38)(10,-3.74)
\pscustom[]{\psline(5.01,1.24)(15,-8.74)
\psbezier(15,-8.74)(15,-8.74)(15,-8.74)
\psline(15,-8.74)(15,0)
\psline(15,0)(5.01,10)
\psline(5.01,10)(5.01,1.24)
\psbezier(5.01,1.24)(5.01,1.24)(5.01,1.24)
\psline(5.01,1.24)(5.01,8.12)
}
\rput(9.38,0.62){$\shh$}
\psline{->}(10,5)(10,9.38)
\rput(10,11.88){$\dh$}
\end{pspicture}
 
 & = & \def\JPicScale{.55}\renewcommand{\dh}{\Nn}
\newcommand{\ahh}{\scriptstyle n}\newcommand{\shh}{\scriptstyle \TK}\newcommand{\Lhh}{\MMmm} 
\ifx\JPicScale\undefined\def\JPicScale{1}\fi
\psset{unit=\JPicScale mm}
\psset{linewidth=0.3,dotsep=1,hatchwidth=0.3,hatchsep=1.5,shadowsize=1,dimen=middle}
\psset{dotsize=0.7 2.5,dotscale=1 1,fillcolor=black}
\psset{arrowsize=1 2,arrowlength=1,arrowinset=0.25,tbarsize=0.7 5,bracketlength=0.15,rbracketlength=0.15}
\begin{pspicture}(0,0)(21.17,11.88)
\rput(7.5,-13.12){$\dh$}
\rput(18.75,-13.12){$\ahh$}
\psline{->}(7.5,-10)(7.5,-1.25)
\rput(20,-4.38){$\Lhh$}
\rput(8.75,1.88){$\shh$}
\psline{->}(10,5)(10,9.38)
\rput(10,11.88){$\dh$}
\psline(5,10)
(15.62,-0.62)
(11.25,-5)
(5,1.25)(5,10)
\rput{0}(18.75,-13.12){\psellipse[](0,0)(2.42,-2.42)}
\pscustom[]{\psline{<-}(13.75,-3.12)(18.75,-8.12)
\psbezier(18.75,-8.12)(18.75,-8.12)(18.75,-8.12)
\psline{-}(18.75,-8.12)(18.75,-10.62)
}
\end{pspicture}
 
\eea
\vspace{.5\baselineskip}

\noindent The normalization now implies that the grading of the universal evaluators is similarly internalized. Note, however, that this does not imply that every sequence of computable functions $<f_n> \in \prod_{n\lt \infty} \CCC_n(A,B)$ comes from some $f\in \CCC(A\otimes \MMmm, B)$. This would imply that a computable limit $f_\infty \in\CCC(A,B)$ with $(f_\infty)\restr_n = f_n$ always exists, which is not the case, as many uncomputable functions have computable approximations \cite{GoldE:limiting}. Kolmogorov complexity provides the basic examples.

\subsection{From Blum measures to normal measures}
 \be{defn}\label{def-blum}
A \emph{Blum measure}\/ (or a \emph{step-counting function} \cite{BlumM:axioms}) is a computation $\CX^A \in \CCC(\PP A ,\MMmm)$ where 
\begin{enumerate}[(i)]
\item\label{cond-halting} $I\tto{F\otimes a} \PP \otimes A \tto{\CX^A}\MMmm$ halts if and only if \\
$I\tto{F\otimes a} \PP \otimes A \tto{\UK^{AB}} B$ halts

\item\label{cond-decidable} $\PP \otimes A \otimes \MMmm \tto{\CX^A\otimes \MMmm} \MMmm \otimes \MMmm \tto \olessthan \Bits$ is decidable.
\end{enumerate}
\ee{defn}
While condition (\ref{cond-halting}) says that the abstract complexity measures are closely related with the universal evaluators, condition (\ref{cond-decidable}) says that they are essentially different. The following definition attempts to capture that difference.

\be{defn}
A \emph{notion of complexity}\/ 
is a triple $(\GK^\ast, \GK_\ast, \olessthan)$ where 
$\GK^\ast:\PP \rightleftarrows \MMmm : \GK_\ast$ 
 are halting functions\footnote{The standard terminology in recursion theory is \emph{"total recursive functions"}. But in other parts of mathematics, a function is always total on its domain, unless we don't know the domain, and it is specified that it is a partial. And moreover, a Turning machine may  write some data on the output tape, and thus provide an output, without halting. So it seems necessary to specify that it is a halting function, and unnecessary to specify that it is total.} in  $\CCC_\natural$,  the predicate $\PP\otimes \PP\tto{\olessthan}\Bits$ is in $\CCC^\natural_0$, and they together satisfy
\bea\label{eq-notion}
\def\JPicScale{.5}\newcommand{\Fhh}{\PP}\renewcommand{\dh}{\PP}\newcommand{\EQ}{=}\newcommand{\trth}{\scriptstyle 1}\newcommand{\ahh}{\scriptscriptstyle\GK_\ast}\newcommand{\order}{\scriptstyle \olessthan}\newcommand{\Tval}{\Bits}\newcommand{\bhh}{A}\newcommand{\Lhh}{\MMmm}\newcommand{\qhh}{\scriptstyle\TK} \newcommand{\shh}{\scriptstyle\SK^A} \newcommand{\thh}{\scriptstyle \TK_\infty}
\ifx\JPicScale\undefined\def\JPicScale{1}\fi
\psset{unit=\JPicScale mm}
\psset{linewidth=0.3,dotsep=1,hatchwidth=0.3,hatchsep=1.5,shadowsize=1,dimen=middle}
\psset{dotsize=0.7 2.5,dotscale=1 1,fillcolor=black}
\psset{arrowsize=1 2,arrowlength=1,arrowinset=0.25,tbarsize=0.7 5,bracketlength=0.15,rbracketlength=0.15}
\begin{pspicture}(0,0)(148.12,29.38)
\rput(25.62,1.88){$\ahh$}
\rput(19.38,-20){$\Lhh$}
\psline(19.38,4.38)
(28.12,4.38)
(28.12,-4.38)
(27.49,-3.77)(19.38,4.38)
\rput{0}(16.88,14.38){\psellipse[](0,0)(3.44,-3.44)}
\psline{<-}(19.38,11.87)
(23.75,7.5)(23.75,4.38)
\rput(16.88,14.37){$\order$}
\psline{->}(16.88,18.13)(16.88,22.5)
\rput(16.88,25){$\Tval$}
\psline{<-}(14.38,11.89)
(10,7.5)(10,3.12)
\psline{<-}(23.76,-0)
(23.75,-6.25)(19.38,-10.62)
\rput{0}(19.38,-10.62){\psellipse[fillstyle=solid](0,0)(1.48,-1.48)}
\psline{->}(19.38,-16.88)(19.38,-11.87)
\rput(7.5,-20){$\dh$}
\psline{->}(7.5,-16.88)(7.5,-3.12)
\psline(5,8.12)
(15.62,-2.5)
(11.25,-6.88)
(5,-0.62)(5,8.12)
\pscustom[]{\psline{<-}(13.75,-5)(18.75,-10)
\psbezier(18.75,-10)(18.75,-10)(18.75,-10)
\psline{-}(18.75,-10)(19.38,-10.62)
}
\rput(35,0){$\EQ$}
\psline{->}(43.12,-16.88)(43.12,-2.5)
\rput{0}(48.12,6.88){\psellipse[](0,0)(2.97,-2.97)}
\psline{->}(48.12,10)(48.12,21.88)
\rput(48.12,25){$\Tval$}
\rput{0}(43.12,-2.5){\psellipse[fillstyle=solid](0,0)(1.48,-1.48)}
\rput{0}(53.12,-2.5){\psellipse[fillstyle=solid](0,0)(1.48,-1.48)}
\psline{->}(53.12,-16.88)(53.12,-2.5)
\newrgbcolor{userFillColour}{0.2 0.2 0.2}
\rput(48.12,6.88){$\trth$}
\rput(53.12,-20){$\Lhh$}
\rput(43.12,-20){$\dh$}
\rput(94.37,10){$\ahh$}
\psline{->}(91.87,-10.62)(91.87,-6.88)
\psline{->}(93.12,11.89)(97.5,16.24)
\rput(106.25,-23.12){$\Lhh$}
\pscustom[]{\psline(88.13,-3.12)(96.25,-11.24)
\psbezier(96.25,-11.24)(96.25,-11.24)(96.25,-11.24)
\psline(96.25,-11.24)(96.25,-0.62)
\psline(96.25,-0.62)(88.13,7.51)
\psline(88.13,7.51)(88.13,-3.12)
\psbezier(88.13,-3.12)(88.13,-3.12)(88.13,-3.12)
\psline(88.13,-3.12)(88.13,4.38)
}
\psline(88.13,11.88)
(96.87,11.88)
(96.87,3.12)
(96.25,3.72)(88.13,11.88)
\psline{->}(92.5,3.12)(92.5,7.51)
\psline{->}(81.88,-19.38)(81.88,-10.62)
\psline{->}(90,-19.24)(90,-15.01)
\rput(90,-23.12){$\bhh$}
\rput(87.5,-10.62){$\shh$}
\psline(78.12,-6.88)
(86.25,-15)
(96.25,-15.01)
(87.5,-6.25)
(87.5,-6.25)
(77.5,-6.25)
(78.12,-6.88)(77.5,-6.25)
\rput(92.5,-1.26){$\thh$}
\rput(81.88,-23.12){$\Fhh$}
\rput{0}(100,18.75){\psellipse[](0,0)(3.44,-3.43)}
\psline{<-}(102.5,16.24)
(106.25,12.5)(106.25,-18.76)
\rput(100,18.74){$\order$}
\psline{->}(100,22.5)(100,26.88)
\rput(100,29.38){$\Tval$}
\rput(113.75,0){$\EQ$}
\rput(142.5,1.24){$\ahh$}
\psline{->}(128.12,-10.62)(128.12,-6.88)
\rput(143.75,-23.12){$\Lhh$}
\pscustom[]{\psline(124.38,-3.12)(132.5,-11.24)
\psbezier(132.5,-11.24)(132.5,-11.24)(132.5,-11.24)
\psline(132.5,-11.24)(132.5,-0.62)
\psline(132.5,-0.62)(124.38,7.51)
\psline(124.38,7.51)(124.38,-3.12)
\psbezier(124.38,-3.12)(124.38,-3.12)(124.38,-3.12)
\psline(124.38,-3.12)(124.38,4.38)
}
\psline(148.12,-1.26)
(139.37,-1.26)
(139.37,7.49)
(140,6.9)(148.12,-1.26)
\psline{->}(143.75,-19.38)(143.75,-1.26)
\psline{->}(118.75,-20)(118.75,-11.25)
\psline{->}(126.25,-19.24)(126.25,-15.01)
\rput(126.25,-23.12){$\bhh$}
\rput(123.75,-10.62){$\shh$}
\pscustom[]{\psline(113.75,-6.25)(122.5,-15)
\psline(122.5,-15)(132.5,-15.01)
\psline(132.5,-15.01)(123.75,-6.25)
\psline(123.75,-6.25)(123.75,-6.25)
\psline(123.75,-6.25)(113.75,-6.25)
\psline(113.75,-6.25)(114.38,-6.25)
\psbezier(114.38,-6.25)(114.38,-6.25)(114.38,-6.25)
}
\rput(128.75,-1.26){$\thh$}
\rput(118.75,-23.12){$\Fhh$}
\rput{0}(136.25,18.75){\psellipse[](0,0)(3.44,-3.43)}
\psline{<-}(133.75,16.24)
(128.75,11.24)(128.75,3.12)
\rput(136.25,18.74){$\order$}
\psline{->}(136.25,22.5)(136.25,26.88)
\rput(136.25,29.38){$\Tval$}
\rput(130.62,9.38){}
\psline{<-}(138.75,16.24)
(143.75,11.24)(143.75,3.12)
\rput(9.38,-0.62){$\qhh$}
\end{pspicture}
 
\eea
\vspace{2.5\baselineskip}
\ee{defn}

\be{defn}
A \emph{normal complexity measure}\/ in a graded monoidal computer is a family of computations $\CX^A\in \CCC(\PP A, \MMmm)$ for all $A$ in $\CCC$ for which there is a notion of complexity $(\GK^\ast, \GK_\ast, \olessthan)$ that normalizes it:
\bea\label{eq-cx-decomp}
\def\JPicScale{.6}\renewcommand{\dh}{\Nn}
\newcommand{\bhh}{A} \newcommand{\Lhh}{\MMmm} \newcommand{\ahh}{\scriptstyle \CX^{AB}}  & = & \def\JPicScale{.6}\newcommand{\Fhh}{\PP}\newcommand{\ahh}{\scriptscriptstyle\GK^\ast}\newcommand{\bhh}{A}\newcommand{\Lhh}{\MMmm}\newcommand{\shh}{\scriptstyle\SK^A} \newcommand{\thh}{\scriptstyle \TK_\infty}
\eea
\vspace{2\baselineskip}
\ee{defn}

\be{prop}
A normal complexity measure is a Blum measure where evaluating the complexity of a program is not substantially harder than evaluating the program itself. More precisely, there is $\chi\in \CCC_\natural(\MMmm,\MMmm)$ and a program $C$ for all $\CX$ so that

\pagebreak
\bea\label{eq-cx-norm}
\def\JPicScale{.5}\newcommand{\Fhh}{\PP}\newcommand{\ahh}{\scriptscriptstyle \chi}\newcommand{\order}{\scriptstyle \olessthan}\newcommand{\Tval}{\Bits}\newcommand{\bhh}{A}\newcommand{\Lhh}{\MMmm}\newcommand{\chh}{\scriptstyle\CX^{\PP A}} \newcommand{\chhh}{\scriptstyle{\CX^A}} \newcommand{\thh}{\scriptscriptstyle C}
\ifx\JPicScale\undefined\def\JPicScale{1}\fi
\psset{unit=\JPicScale mm}
\psset{linewidth=0.3,dotsep=1,hatchwidth=0.3,hatchsep=1.5,shadowsize=1,dimen=middle}
\psset{dotsize=0.7 2.5,dotscale=1 1,fillcolor=black}
\psset{arrowsize=1 2,arrowlength=1,arrowinset=0.25,tbarsize=0.7 5,bracketlength=0.15,rbracketlength=0.15}
\begin{pspicture}(0,0)(46.88,50.62)
\rput(41.88,17.5){$\ahh$}
\psline(36.26,20.02)
(44.38,20.02)
(44.38,11.9)
(44.38,11.88)(36.26,20.02)
\psline{->}(40,8.12)(40,15.62)
\psline{->}(18.12,-35.62)(18.12,-26.88)
\rput(28.12,-39.38){$\bhh$}
\rput(0.62,-9.38){$\thh$}
\rput(16.88,-39.38){$\Fhh$}
\rput{0}(25.62,40){\psellipse[](0,0)(3.44,-3.44)}
\psline{<-}(28.12,37.5)
(40,25.62)(40,20)
\rput(25.62,40){$\order$}
\psline{->}(25.62,43.76)(25.62,48.12)
\rput(25.62,50.62){$\Tval$}
\pscustom[]{\psline(6.88,-4.38)(16.88,-4.38)
\psbezier(16.88,-4.38)(16.88,-4.38)(16.88,-4.38)
\psline(16.88,-4.38)(16.88,8.12)
\psline(16.88,8.12)(3.12,8.12)
\psline(3.12,8.12)(3.12,3.75)
\psbezier(3.12,3.75)(3.12,3.75)(3.12,3.75)
\psbezier(3.12,3.75)(3.12,3.75)(3.12,3.75)
}
\pscustom[]{\psline(-1.88,3.75)(6.25,-4.38)
\psline(6.25,-4.38)(16.88,-4.38)
\psline(16.88,-4.38)(14.38,-4.38)
\psline(14.38,-4.38)(6.25,-4.38)
\psline(6.25,-4.38)(-1.88,3.75)
\psbezier(-1.88,3.75)(-1.88,3.75)(-1.88,3.75)
\psline(-1.88,3.75)(3.12,3.75)
}
\psline{<-}(23.12,37.52)
(10,24.38)(10,8.12)
\psline{<-}(8.75,-4.99)
(8.75,-16.25)(18.12,-25.62)
\psline{<-}(31.25,0.62)
(31.25,-12.5)(17.5,-26.25)
\psline[border=1.05]{<-}(14.38,-5)
(14.38,-14.38)(29.38,-29.38)
\psline{<-}(41.88,-4.38)
(41.88,-15.62)(28.12,-29.38)
\rput{0}(18.12,-25.62){\psellipse[fillstyle=solid](0,0)(1.48,-1.48)}
\rput{0}(28.12,-29.38){\psellipse[fillstyle=solid](0,0)(1.48,-1.48)}
\psline{->}(28.12,-36.25)(28.12,-30.62)
\newrgbcolor{userFillColour}{0.2 0.2 0.2}
\rput(10,1.88){$\chh$}
\pscustom[]{\psline(36.88,-4.38)(46.88,-4.38)
\psbezier(46.88,-4.38)(46.88,-4.38)(46.88,-4.38)
\psline(46.88,-4.38)(46.88,8.12)
\psline(46.88,8.12)(33.12,8.12)
\psline(33.12,8.12)(33.12,3.75)
\psbezier(33.12,3.75)(33.12,3.75)(33.12,3.75)
\psbezier(33.12,3.75)(33.12,3.75)(33.12,3.75)
}
\pscustom[]{\psline(28.12,3.75)(36.25,-4.38)
\psline(36.25,-4.38)(46.88,-4.38)
\psline(46.88,-4.38)(44.38,-4.38)
\psline(44.38,-4.38)(36.25,-4.38)
\psline(36.25,-4.38)(28.12,3.75)
\psbezier(28.12,3.75)(28.12,3.75)(28.12,3.75)
\psline(28.12,3.75)(33.12,3.75)
}
\psline(6.25,-11.9)
(-1.88,-11.9)
(-1.88,-3.78)
(-1.88,-3.75)(6.25,-11.9)
\psline{->}(1.88,-7.5)(1.88,0)
\rput(40,1.88){$\chhh$}
\end{pspicture}
  & = & \def\JPicScale{.5}\newcommand{\Fhh}{\PP}\newcommand{\trth}{\scriptstyle 1}\newcommand{\Tval}{\Bits}\newcommand{\bhh}{A}\newcommand{\Lhh}{\MMmm}\newcommand{\shh}{\scriptstyle\SK^A} \newcommand{\thh}{\scriptstyle \TK_\infty}
\ifx\JPicScale\undefined\def\JPicScale{1}\fi
\psset{unit=\JPicScale mm}
\psset{linewidth=0.3,dotsep=1,hatchwidth=0.3,hatchsep=1.5,shadowsize=1,dimen=middle}
\psset{dotsize=0.7 2.5,dotscale=1 1,fillcolor=black}
\psset{arrowsize=1 2,arrowlength=1,arrowinset=0.25,tbarsize=0.7 5,bracketlength=0.15,rbracketlength=0.15}
\begin{pspicture}(0,0)(16.48,21.88)
\psline{->}(5,-18.75)(5,-4.38)
\rput(15,-23.75){$\bhh$}
\rput(5,-23.75){$\Fhh$}
\rput{0}(10,5.62){\psellipse[](0,0)(2.97,-2.97)}
\psline{->}(10,8.75)(10,16.88)
\rput(10,21.88){$\Tval$}
\rput{0}(5,-4.38){\psellipse[fillstyle=solid](0,0)(1.48,-1.48)}
\rput{0}(15,-4.38){\psellipse[fillstyle=solid](0,0)(1.48,-1.48)}
\psline{->}(15,-18.75)(15,-4.38)
\newrgbcolor{userFillColour}{0.2 0.2 0.2}
\rput(10,5.62){$\trth$}
\end{pspicture}
\eea
\vspace{3\baselineskip}
\ee{prop}

\section{Summary}\label{Sec-Outro}
Theoretical computer science works with a wide gamut of different models of computation. While they all implement the same family of computable functions, they also confront us with a wide gamut of different low-level machine languages that we use to describe computations. The fact that all these different machines compute the same functions through radically different computational processes is usually celebrated as a conceptual miracle, giving rise to Church's Thesis. But the conceptual miracle gives way to the technical difficulties when it comes to programming in these machine languages. 
Proving that a feasible computation in one model will not become unfeasible in another model is generally not an easy task. With the questions that involve translating a complexity concept from one model to another often beyond reach, it seems fair to say that complexity theory is largely a mosaic of machine dependent concepts. The programming routines of \emph{hiding the implementation details} and of  \emph{high-level languages}, on which the practice of computation has been based for more than 50 years, have not yet reached the theory of computation.

A notable early effort towards machine independent complexity theory was initiated by Blum \cite{BlumM:axioms,BlumM:speedup}, and pursued by others \cite{MeyerA:speedup,LevinL:fundamental}. We implement and investigate some of their ideas in the framework of monoidal computer. Normal complexity measures, that admit a version of Kleene's normal form, emerge as an interesting concept. There is a sense in which the space of normal complexity measures is spanned by the time and the space complexity measures, as an orthogonal basis. This will have to be elaborated in the future work, together with all the proofs and details omitted from this extended abstract. While the measures that are not normal provide important design time information about algorithms, and for theoretical analyses, normal measures can also be used at run time, as practical tools of computation, e.g. to set the bounds for hypothesis testing, inductive inference and algorithmic learning.

\bibliographystyle{IEEEtranS}
\bibliography{IEEEabrv,ref-14-LICS,ref-moncomp-1,PavlovicD}

\begin{thebibliography}{10}
\providecommand{\url}[1]{#1}
\csname url@samestyle\endcsname
\providecommand{\newblock}{\relax}
\providecommand{\bibinfo}[2]{#2}
\providecommand{\BIBentrySTDinterwordspacing}{\spaceskip=0pt\relax}
\providecommand{\BIBentryALTinterwordstretchfactor}{4}
\providecommand{\BIBentryALTinterwordspacing}{\spaceskip=\fontdimen2\font plus
\BIBentryALTinterwordstretchfactor\fontdimen3\font minus
  \fontdimen4\font\relax}
\providecommand{\BIBforeignlanguage}[2]{{%
\expandafter\ifx\csname l@#1\endcsname\relax
\typeout{** WARNING: IEEEtranS.bst: No hyphenation pattern has been}%
\typeout{** loaded for the language `#1'. Using the pattern for}%
\typeout{** the default language instead.}%
\else
\language=\csname l@#1\endcsname
\fi
#2}}
\providecommand{\BIBdecl}{\relax}
\BIBdecl

\bibitem{AbramskyS:retracing}
S.~Abramsky, ``Retracing some paths in process algebra,'' in \emph{CONCUR},
  ser. Lecture Notes in Computer Science, U.~Montanari and V.~Sassone, Eds.,
  vol. 1119.\hskip 1em plus 0.5em minus 0.4em\relax Springer, 1996, pp. 1--17.

\bibitem{AspertiA:rice}
A.~Asperti, ``The intensional content of {Rice's Theorem},'' in
  \emph{Proceedings of the 35th Annual ACM SIGPLAN-SIGACT Symposium on
  Principles of Programming Languages}, ser. POPL '08.\hskip 1em plus 0.5em
  minus 0.4em\relax New York, NY, USA: ACM, 2008, pp. 113--119.

\bibitem{BarthaM:monoidal-TM}
M.~{Bartha}, ``\BIBforeignlanguage{English}{{The monoidal structure of Turing
  machines}},'' \emph{\BIBforeignlanguage{English}{{Math. Struct. Comput.
  Sci.}}}, vol.~23, no.~2, pp. 204--246, 2013.

\bibitem{Bernstein-Vazirani}
E.~Bernstein and U.~V. Vazirani, ``Quantum complexity theory,'' \emph{SIAM J.
  Comput.}, vol.~26, no.~5, pp. 1411--1473, 1997.

\bibitem{BirkedalL}
L.~Birkedal, ``A general notion of realizability,'' \emph{Bulletin of Symbolic
  Logic}, vol.~8, no.~2, pp. 266--282, 2002.

\bibitem{BlumM:axioms}
\BIBentryALTinterwordspacing
M.~Blum, ``A machine-independent theory of the complexity of recursive
  functions,'' \emph{J. ACM}, vol.~14, no.~2, pp. 322--336, Apr. 1967.
  [Online]. Available: \url{http://doi.acm.org/10.1145/321386.321395}
\BIBentrySTDinterwordspacing

\bibitem{BlumM:speedup}
------, ``On effective procedures for speeding up algorithms,'' in \emph{STOC},
  P.~C. Fischer, S.~Ginsburg, and M.~A. Harrison, Eds.\hskip 1em plus 0.5em
  minus 0.4em\relax ACM, 1969, pp. 43--53.

\bibitem{Carboni-Walters}
A.~Carboni and R.~F. Walters, ``Cartesian bicategories, {I},'' \emph{J. of Pure
  and Applied Algebra}, vol.~49, pp. 11--32, 1987.

\bibitem{Cockett-Hofstra}
J.~R.~B. Cockett and P.~J.~W. Hofstra, ``Introduction to {Turing} categories,''
  \emph{Ann. Pure Appl. Logic}, vol. 156, no. 2-3, pp. 183--209, 2008.

\bibitem{PavlovicD:CQstruct}
B.~Coecke, \'{E}ric Oliver~Paquette, and {Dusko Pavlovic}, ``Classical and
  quantum structuralism,'' in \emph{Semantical Techniques in Quantum
  Computation}, S.~Gay and I.~Mackie, Eds.\hskip 1em plus 0.5em minus
  0.4em\relax Cambridge University Press, 2009, pp. 29--69.

\bibitem{PavlovicD:QMWS}
B.~Coecke and D.~Pavlovic, ``Quantum measurements without sums,'' in
  \emph{Mathematics of Quantum Computing and Technology}, G.~Chen, L.~Kauffman,
  and S.~Lamonaco, Eds.\hskip 1em plus 0.5em minus 0.4em\relax Taylor and
  Francis, 2007, arxiv.org/quant-ph/0608035.

\bibitem{eilenberg-elgot}
S.~Eilenberg and C.~Elgot, \emph{Recursiveness}, ser. ACM Monograph.\hskip 1em
  plus 0.5em minus 0.4em\relax Academic Press, 1970.

\bibitem{MisloveM:compendium}
G.~Gierz, K.~H. Hofmann, K.~Keimel, J.~D. Lawson, and M.~W. Mislove, \emph{A
  compendium of continous lattices}.\hskip 1em plus 0.5em minus 0.4em\relax
  Cambridge University Press, 2003, (First edition 1980).

\bibitem{GoldE:limiting}
E.~M. Gold, ``Limiting recursion,'' \emph{J. Symb. Log.}, vol.~30, no.~1, pp.
  28--48, 1965.

\bibitem{HofstraW13}
P.~J.~W. Hofstra and M.~A. Warren, ``Combinatorial realizability models of type
  theory,'' \emph{Ann. Pure Appl. Logic}, vol. 164, no.~10, pp. 957--988, 2013.

\bibitem{HylandJ:efft}
J.~Hyland, ``The effective topos,'' in \emph{The {L.E.J. Brouwer} Centenary
  Symposium}, A.~Troelstra and D.~V. Dalen, Eds.\hskip 1em plus 0.5em minus
  0.4em\relax North Holland Publishing Company, 1982, pp. 165--216.

\bibitem{Joyal-Street:geometry}
A.~Joyal and R.~Street, ``The geometry of tensor calculus {I},'' \emph{Adv. in
  Math.}, vol.~88, pp. 55--113, 1991.

\bibitem{KellyM:Enriched}
G.~M. Kelly, \emph{\BIBforeignlanguage{English}{Basic concepts of enriched
  category theory}}.\hskip 1em plus 0.5em minus 0.4em\relax Cambridge
  University Press, 1982.

\bibitem{KleeneSC1936}
S.~C. Kleene, ``\BIBforeignlanguage{English}{{General recursive functions of
  natural numbers.}}'' \emph{\BIBforeignlanguage{English}{{Math. Ann. 112,
  727-742}}}, 1936.

\bibitem{Lambek-Scott:book}
J.~Lambek and P.~J. Scott, \emph{{Introduction to Higher-Order Categorical
  Logic}}.\hskip 1em plus 0.5em minus 0.4em\relax {Cambridge University Press},
  {1986}.

\bibitem{LawvereFW:adjf}
F.~W. Lawvere, ``Adjointness in foundations,'' \emph{Dialectica}, vol.~23, pp.
  281--296, 1969.

\bibitem{CatSOS}
M.~Lenisa, J.~Power, and H.~Watanabe, ``Category theory for operational
  semantics,'' \emph{Theor. Comput. Sci.}, vol. 327, no. 1-2, pp. 135--154,
  2004.

\bibitem{LevinL:fundamental}
L.~A. Levin, ``Computational complexity of functions,'' \emph{Theoretical
  Computer Science}, vol. 157, no.~2, pp. 267 -- 271, 1996.

\bibitem{MacLaneS:CWM}
S.~MacLane, \emph{Categories for the Working Mathematician}, ser. Graduate
  Texts in Mathematics.\hskip 1em plus 0.5em minus 0.4em\relax Springer-Verlag,
  1971, no.~5.

\bibitem{MeyerA:speedup}
A.~R. Meyer and P.~C. Fischer, ``Computational speed-up by effective
  operators,'' \emph{J. Symb. Log.}, vol.~37, no.~1, pp. 55--68, 1972.

\bibitem{heller-dipaola}
R.~A.~D. Paola and A.~Heller, ``{Dominical Categories: Recursion Theory without
  Elements},'' \emph{J. Symbolic Logic}, vol.~52, no.~3, pp. 594--635, 1987.

\bibitem{PavlovicD:MSCS97}
D.~Pavlovic, ``Categorical logic of names and abstraction in action calculus,''
  \emph{Math. Structures in Comp. Sci.}, vol.~7, pp. 619--637, 1997.

\bibitem{PavlovicD:QI09}
------, ``Quantum and classical structures in nondeterministic computation,''
  in \emph{Proceedings of Quantum Interaction 2009}, ser. Lecture Notes in
  Artificial Intelligence, P.~Bruza, D.~Sofge, and K.~{van Rijsbergen}, Eds.,
  vol. 5494.\hskip 1em plus 0.5em minus 0.4em\relax Springer Verlag, 2009, pp.
  143--158, arxiv.org:0812.2266.

\bibitem{PavlovicD:NSPW11}
------, ``Gaming security by obscurity,'' in \emph{Proceedings of NSPW 2011},
  C.~Gates and C.~Hearley, Eds.\hskip 1em plus 0.5em minus 0.4em\relax New
  York, NY, USA: ACM, 2011, pp. 125--140, arxiv:1109.5542.

\bibitem{PavlovicD:QPL09}
------, ``Relating toy models of quantum computation: comprehension,
  complementarity and dagger autonomous categories,'' \emph{E. Notes in Theor.
  Comp. Sci.}, vol. 270, no.~2, pp. 121--139, 2011, arxiv.org:1006.1011.

\bibitem{PavlovicD:Qabs12}
\BIBentryALTinterwordspacing
------, ``Geometry of abstraction in quantum computation,'' \emph{Proceedings
  of Symposia in Applied Mathematics}, vol.~71, pp. 233--267, 2012,
  arxiv.org:1006.1010. [Online]. Available:
  \url{http://www.comlab.ox.ac.uk//files/2533/RR-09-13.pdf}
\BIBentrySTDinterwordspacing

\bibitem{PavlovicD:IC12}
------, ``Monoidal computer {I}: {Basic computability by string diagrams},''
  \emph{Information and Computation}, vol. 226, pp. 94--116, 2013,
  arxiv:1208.5205.

\bibitem{PavlovicD:CTCS97}
D.~Pavlovic and S.~Abramsky, ``Specifying interaction categories,'' in
  \emph{Category Theory and Computer Science '97}, ser. Lecture Notes in
  Computer Science, E.~Moggi and G.~Rosolini, Eds., vol. 1290.\hskip 1em plus
  0.5em minus 0.4em\relax Springer Verlag, 1997, pp. 147--158.

\bibitem{Penrose}
R.~Penrose, ``{Structure of space-time},'' in \emph{Batelle Rencontres, 1967},
  C.~DeWitt and J.~Wheeler, Eds.\hskip 1em plus 0.5em minus 0.4em\relax
  Benjamin, 1968.

\bibitem{RogersH:book}
H.~Rogers, Jr., \emph{Theory of recursive functions and effective
  computability}.\hskip 1em plus 0.5em minus 0.4em\relax Cambridge, MA, USA:
  MIT Press, 1987.

\bibitem{SchutzenbergerM:65}
M.~P. Sch{\"u}tzenberger, ``On finite monoids having only trivial subgroups,''
  \emph{Information and Control}, vol.~8, no.~2, pp. 190--194, 1965.

\bibitem{ScottD:domains}
D.~S. Scott, ``Domains for denotational semantics,'' in \emph{Automata,
  Languages and Programming}, ser. Lecture Notes in Computer Science,
  M.~Nielsen and E.~M. Schmidt, Eds.\hskip 1em plus 0.5em minus 0.4em\relax
  Springer, 1982, vol. 140, pp. 577--610.

\bibitem{selinger2011survey}
P.~Selinger, ``A survey of graphical languages for monoidal categories,'' in
  \emph{New structures for physics}.\hskip 1em plus 0.5em minus 0.4em\relax
  Springer, 2011, pp. 289--355.

\bibitem{Stoy}
J.~E. Stoy, \emph{Denotational Semantics: The Scott-Strachey Approach to
  Programming Language Theory}.\hskip 1em plus 0.5em minus 0.4em\relax
  Cambridge, MA, USA: MIT Press, 1977.

\bibitem{ThompsonS:book}
S.~Thompson, \emph{Haskell - the craft of functional programming}, ser.
  International computer science series.\hskip 1em plus 0.5em minus 0.4em\relax
  Addison-Wesley, 1996.

\end{thebibliography}
%



\appendix

\section*{Proof sketch for Prop.~\ref{prop-us}}
Given $\gamma$ as in Def.~\ref{def:moncomp}(\ref{itemgamma}), the universal evaluators $\UK$ are defined
\bea\label{uone}
\UK^{AB} = \gamma^{AB}_\PP(\id_\PP) 
\eea
and then the partial evaluators $\SK$ are determined by
\bea\label{sone}
\gamma_{\PP\otimes A}^{BC}\left(\SK^{(AB)C}\right) & = & \UK^{(AB)C}
\eea
The other way around, if the universal evaluators $\UK$ are given as in Prop.~\ref{prop-us}(\ref{itemuniv}), then the natural transformation $\gamma^{AB}$  as in Def.~\ref{def:moncomp}(\ref{itemgamma}) can be defined by
\bea\label{gammadef}
\gamma^{AB}_X\left(X\tto h \PP\right)\ = \ XA\tto{h  A} \PP A \tto{\UK^{AB}} B
\eea
This is surjective because
\begin{itemize}
\item \ref{prop-us}(\ref{itemuniv}) says that every $g\in \CCC(XA, B)$ decomposes to 
\bear
XA = IXA \tto{\widetilde g XA} \PP XA \tto{\UK^{(XA)B}} B
\eear
for some  $\widetilde g \in \CCC^\natural(I,\PP)$, whereas
\item  \ref{prop-us}(\ref{itempart}) and \eqref{gammadef} give 
\bear
\gamma^{AB}_X\left(IX \tto{\widetilde g X} \PP X \tto {\SK^{(XA)B}} \PP\right) & = &  g
\eear
\end{itemize}

\section*{Proof sketch for Prop.~\ref{prop-ustw}}
Given the natural transformations as in Def.~\ref{def-gmc}, the evaluators from Prop~\ref{prop-ustw} are defined by
\bea
\UK_n^{AB} & = & \gamma^{AB}_\PP\left(\id_\PP\right)\restr_n \label{eq:UK}\\
\SK^A & = & \sigma^A_{\PP}\left(\id_\PP\right)\label{eq:SK}\\
\TK_n & = & \tau^I_\PP \left(\id_\PP\right) \restr_n \label{eq:TK}\\
\VK^B & = & \vartheta^{B}_{\PP}\left(\id_\PP\right)\label{eq:VK}
\eea
It is not hard to check that the conditions of Def.~\ref{def-gmc} are equivalent with the conditions of Prop.~\ref{prop-ustw}.

\end{document}